\documentclass[letterpaper, 10 pt, conference]{ieeeconf}  

\IEEEoverridecommandlockouts                              

\title{\LARGE \bf
	The optimal sequence for reset controllers
}

\author{Chengwei Cai$^{1}$, Ali Ahmadi Dastjerdi$^{1}$, Niranjan Saikumar$^{1}$ and S.H. HosseinNia$^{1}$
	\thanks{$^{1}$C. Cai, A. Ahmadi Dastjerdi, N.saikumar and S.H HosseinNia are with the Faculty of Department of Precision and
		Microsystems Engineering, Delft University of Technology, Delft, The Netherlands
		{\tt\small C.Cai-2@student.tudelft.nl, A.AhmadiDastjerdi@tudelft.nl, N.saikumar@tudelft.nl, and\newline s.h.hosseinniakani@tudelft.nl}}%
}

\usepackage{multirow}
\usepackage{graphicx}
\usepackage{cancel} 
\usepackage{amsmath,bm}
\usepackage{tikz}
\usepackage{mathdots}
\usepackage{yhmath}
\usepackage{cite}
\usepackage{cancel}
\usepackage{color}
\usepackage{siunitx}
\usepackage{array}
\usepackage{multirow}
\usepackage{amssymb}
\usepackage{gensymb}
\usepackage{tabularx}
\usepackage{booktabs}
\usepackage{subcaption}
\usetikzlibrary{fadings}
\usepackage{environ}
   
\begin{document}

	\maketitle
	\thispagestyle{empty}
	\pagestyle{empty}

	\begin{abstract}
		
		PID controllers cannot satisfy the high-performance requirements since they are restricted by the water-bed effect. Thus, the need for a better alternative to linear PID controllers increases due to the rising demands of the high-tech industry. This has led many researchers to explore nonlinear controllers like reset control. Although reset controllers have been widely used to overcome the limitations of linear controllers in literature, the performance of the system varies depending on the relative sequence of controller linear and nonlinear parts. In this paper, the optimal sequence is found using high order sinusoidal input describing functions (HOSIDF). By arranging controller parts according to this strategy, better performance in the sense of precision and control input is achieved. The performance of the proposed sequence is validated on a precision positioning setup. The experimental results demonstrate that the optimal sequence found in theory outperforms other sequences.
	\end{abstract}

	\section{INTRODUCTION}\label{sec:1}
	
	Precision positioning is an important topic in the high-tech industry with applications such as photolithography machines and atomic force microscopes. In these applications, nano-precision controllers with high bandwidth and stability are required to ensure high production quality and speeds. PID controllers, which are one of the most used controllers in the industry owing to their simplicity and ease of tuning, cannot fulfil these control requirements due to their linear nature. This is explained by the water-bed effect which confines the performance of linear controllers so that it is impossible to achieve high bandwidth, stability and precision simultaneously \cite{middleton1991trade,skogestad2007multivariable,schmidt2014design,dastjerdi2019linear}. Reset controllers are a popular nonlinear alternative and have attracted a lot of attention from academia and industry due to their simple structure \cite{clegg1958nonlinear,chen2018beyond,hosseinnia2013basic,marinangeli2018fractional,banos2011reset,valerio2019reset,zhao2019overcoming,zhao2016l2}.
	
	Reset control is a nonlinear control strategy which was introduced in 1958. A traditional reset element resets its state/s to zero when the input signal crosses zero. Clegg introduced the first reset controller by applying reset strategy on a linear integrator \cite{clegg1958nonlinear}. In \cite{horowitz1975non} and \cite{hazeleger2016second}, reset controllers have been extended to First Order Reset Element (FORE) and Second Order Reset Element (SORE) respectively, enabling greater tune-ability and hence applicability in complex systems. Also, several additional strategies have been developed to tune the degree of non-linearity of reset elements such as partial reset and PI+CI \cite{zheng2008development,banos2007definition,barreiro2014reset,zheng2007improved}.
	Reset control has been used to introduce new compensators such as CgLp, CLOC, etc. \cite{palanikumar2018no, saikumar2019constant, saikumar2019resetting, valerio2019reset, saikumar2019complex}.
	
	One of the frequency domain tools for the study of nonlinear controllers is Describing Function (DF), in which the nonlinear controller output is approximated with the first harmonic of Fourier series expansion. Although DF is widely used to analyze and tune reset controllers as well, neglection of the high order harmonics can be seen in the deviation between expected and measured performance \cite{saikumar2019constant, saikumar2019complex}. To investigate the influence of high order harmonics in general nonlinear systems, the concept of high order sinusoidal input describing functions (HOSIDF) was proposed in \cite{nuij2006higher}, which was applied for reset controllers in \cite{heinen2018frequency}.
	
	\begin{figure*}[htbp]
		\center
		\resizebox{\textwidth}{!}{

			\tikzset{every picture/.style={line width=0.75pt}} 
			
			\begin{tikzpicture}[x=0.75pt,y=0.75pt,yscale=-1,xscale=1]
			
			\draw [color={rgb, 255:red, 0; green, 0; blue, 0 }  ,draw opacity=1 ][line width=1.5]    (383,69) -- (564,69) ;
			\draw [shift={(567,69)}, rotate = 180] [fill={rgb, 255:red, 0; green, 0; blue, 0 }  ,fill opacity=1 ][line width=1.5]  [draw opacity=0] (11.61,-5.58) -- (0,0) -- (11.61,5.58) -- cycle    ;
			
			\draw  [color={rgb, 255:red, 74; green, 144; blue, 226 }  ,draw opacity=1 ][line width=2.25]  (718,38) -- (813.5,38) -- (813.5,96) -- (718,96) -- cycle ;
			\draw [color={rgb, 255:red, 0; green, 0; blue, 0 }  ,draw opacity=1 ][line width=1.5]    (961,72) -- (1329,71.01) ;
			\draw [shift={(1332,71)}, rotate = 539.85] [fill={rgb, 255:red, 0; green, 0; blue, 0 }  ,fill opacity=1 ][line width=1.5]  [draw opacity=0] (11.61,-5.58) -- (0,0) -- (11.61,5.58) -- cycle    ;
			
			\draw  [color={rgb, 255:red, 0; green, 0; blue, 0 }  ,draw opacity=1 ][line width=1.5]  (759,171) -- (763.25,171) -- (763.25,135) -- (771.75,135) -- (771.75,171) -- (776,171) -- (767.5,195) -- cycle ;
			\draw  [color={rgb, 255:red, 155; green, 155; blue, 155 }  ,draw opacity=1 ] (428,221) -- (511,221) -- (511,406) -- (428,406) -- cycle ;
			\draw [color={rgb, 255:red, 0; green, 0; blue, 0 }  ,draw opacity=1 ][line width=1.5]    (3.5,312) -- (180,311.02) ;
			\draw [shift={(183,311)}, rotate = 539.6800000000001] [fill={rgb, 255:red, 0; green, 0; blue, 0 }  ,fill opacity=1 ][line width=1.5]  [draw opacity=0] (11.61,-5.58) -- (0,0) -- (11.61,5.58) -- cycle    ;
			
			\draw [color={rgb, 255:red, 155; green, 155; blue, 155 }  ,draw opacity=1 ][line width=1.5]    (512.5,248) -- (689.5,247.02) ;
			\draw [shift={(692.5,247)}, rotate = 539.6800000000001] [fill={rgb, 255:red, 155; green, 155; blue, 155 }  ,fill opacity=1 ][line width=1.5]  [draw opacity=0] (11.61,-5.58) -- (0,0) -- (11.61,5.58) -- cycle    ;
			
			\draw [color={rgb, 255:red, 155; green, 155; blue, 155 }  ,draw opacity=1 ][line width=1.5]    (512.5,288) -- (688.5,288) ;
			\draw [shift={(691.5,288)}, rotate = 180] [fill={rgb, 255:red, 155; green, 155; blue, 155 }  ,fill opacity=1 ][line width=1.5]  [draw opacity=0] (11.61,-5.58) -- (0,0) -- (11.61,5.58) -- cycle    ;
			
			\draw  [color={rgb, 255:red, 155; green, 155; blue, 155 }  ,draw opacity=1 ][fill={rgb, 255:red, 155; green, 155; blue, 155 }  ,fill opacity=1 ] (590,303.5) .. controls (590,302.12) and (591.12,301) .. (592.5,301) .. controls (593.88,301) and (595,302.12) .. (595,303.5) .. controls (595,304.88) and (593.88,306) .. (592.5,306) .. controls (591.12,306) and (590,304.88) .. (590,303.5) -- cycle ;
			\draw  [color={rgb, 255:red, 155; green, 155; blue, 155 }  ,draw opacity=1 ][fill={rgb, 255:red, 155; green, 155; blue, 155 }  ,fill opacity=1 ] (590,345.5) .. controls (590,344.12) and (591.12,343) .. (592.5,343) .. controls (593.88,343) and (595,344.12) .. (595,345.5) .. controls (595,346.88) and (593.88,348) .. (592.5,348) .. controls (591.12,348) and (590,346.88) .. (590,345.5) -- cycle ;
			\draw  [color={rgb, 255:red, 155; green, 155; blue, 155 }  ,draw opacity=1 ][fill={rgb, 255:red, 155; green, 155; blue, 155 }  ,fill opacity=1 ] (590,322.5) .. controls (590,321.12) and (591.12,320) .. (592.5,320) .. controls (593.88,320) and (595,321.12) .. (595,322.5) .. controls (595,323.88) and (593.88,325) .. (592.5,325) .. controls (591.12,325) and (590,323.88) .. (590,322.5) -- cycle ;
			\draw [color={rgb, 255:red, 155; green, 155; blue, 155 }  ,draw opacity=1 ][line width=1.5]    (512.5,383) -- (688.5,383) ;
			\draw [shift={(691.5,383)}, rotate = 180] [fill={rgb, 255:red, 155; green, 155; blue, 155 }  ,fill opacity=1 ][line width=1.5]  [draw opacity=0] (11.61,-5.58) -- (0,0) -- (11.61,5.58) -- cycle    ;
			
			\draw  [color={rgb, 255:red, 155; green, 155; blue, 155 }  ,draw opacity=1 ] (691,229) -- (726,229) -- (726,264) -- (691,264) -- cycle ;
			\draw [color={rgb, 255:red, 155; green, 155; blue, 155 }  ,draw opacity=1 ][line width=1.5]    (727,246) -- (1001,246.99) ;
			\draw [shift={(1004,247)}, rotate = 180.21] [fill={rgb, 255:red, 155; green, 155; blue, 155 }  ,fill opacity=1 ][line width=1.5]  [draw opacity=0] (11.61,-5.58) -- (0,0) -- (11.61,5.58) -- cycle    ;
			
			\draw [color={rgb, 255:red, 155; green, 155; blue, 155 }  ,draw opacity=1 ][line width=1.5]    (727,286) -- (1001,286) ;
			\draw [shift={(1004,286)}, rotate = 180] [fill={rgb, 255:red, 155; green, 155; blue, 155 }  ,fill opacity=1 ][line width=1.5]  [draw opacity=0] (11.61,-5.58) -- (0,0) -- (11.61,5.58) -- cycle    ;
			
			\draw [color={rgb, 255:red, 155; green, 155; blue, 155 }  ,draw opacity=1 ][line width=1.5]    (729,385) -- (1003,385) ;
			\draw [shift={(1006,385)}, rotate = 180] [fill={rgb, 255:red, 155; green, 155; blue, 155 }  ,fill opacity=1 ][line width=1.5]  [draw opacity=0] (11.61,-5.58) -- (0,0) -- (11.61,5.58) -- cycle    ;
			
			\draw  [color={rgb, 255:red, 155; green, 155; blue, 155 }  ,draw opacity=1 ][line width=1.5]  (1327,313.5) .. controls (1327,303.84) and (1334.84,296) .. (1344.5,296) .. controls (1354.16,296) and (1362,303.84) .. (1362,313.5) .. controls (1362,323.16) and (1354.16,331) .. (1344.5,331) .. controls (1334.84,331) and (1327,323.16) .. (1327,313.5) -- cycle ;
			\draw [color={rgb, 255:red, 0; green, 0; blue, 0 }  ,draw opacity=1 ][line width=1.5]    (1362,315) -- (1720,315) ;
			\draw [shift={(1723,315)}, rotate = 180] [fill={rgb, 255:red, 0; green, 0; blue, 0 }  ,fill opacity=1 ][line width=1.5]  [draw opacity=0] (11.61,-5.58) -- (0,0) -- (11.61,5.58) -- cycle    ;
			
			\draw  [color={rgb, 255:red, 74; green, 144; blue, 226 }  ,draw opacity=1 ][dash pattern={on 6.75pt off 4.5pt}][line width=2.25]  (424,214) -- (990,214) -- (990,413) -- (424,413) -- cycle ;
			\draw  [color={rgb, 255:red, 155; green, 155; blue, 155 }  ,draw opacity=1 ][fill={rgb, 255:red, 155; green, 155; blue, 155 }  ,fill opacity=1 ] (706,311.5) .. controls (706,310.12) and (707.12,309) .. (708.5,309) .. controls (709.88,309) and (711,310.12) .. (711,311.5) .. controls (711,312.88) and (709.88,314) .. (708.5,314) .. controls (707.12,314) and (706,312.88) .. (706,311.5) -- cycle ;
			\draw  [color={rgb, 255:red, 155; green, 155; blue, 155 }  ,draw opacity=1 ][fill={rgb, 255:red, 155; green, 155; blue, 155 }  ,fill opacity=1 ] (706,353.5) .. controls (706,352.12) and (707.12,351) .. (708.5,351) .. controls (709.88,351) and (711,352.12) .. (711,353.5) .. controls (711,354.88) and (709.88,356) .. (708.5,356) .. controls (707.12,356) and (706,354.88) .. (706,353.5) -- cycle ;
			\draw  [color={rgb, 255:red, 155; green, 155; blue, 155 }  ,draw opacity=1 ][fill={rgb, 255:red, 155; green, 155; blue, 155 }  ,fill opacity=1 ] (706,330.5) .. controls (706,329.12) and (707.12,328) .. (708.5,328) .. controls (709.88,328) and (711,329.12) .. (711,330.5) .. controls (711,331.88) and (709.88,333) .. (708.5,333) .. controls (707.12,333) and (706,331.88) .. (706,330.5) -- cycle ;
			\draw  [color={rgb, 255:red, 155; green, 155; blue, 155 }  ,draw opacity=1 ][fill={rgb, 255:red, 155; green, 155; blue, 155 }  ,fill opacity=1 ] (870,308.5) .. controls (870,307.12) and (871.12,306) .. (872.5,306) .. controls (873.88,306) and (875,307.12) .. (875,308.5) .. controls (875,309.88) and (873.88,311) .. (872.5,311) .. controls (871.12,311) and (870,309.88) .. (870,308.5) -- cycle ;
			\draw  [color={rgb, 255:red, 155; green, 155; blue, 155 }  ,draw opacity=1 ][fill={rgb, 255:red, 155; green, 155; blue, 155 }  ,fill opacity=1 ] (870,350.5) .. controls (870,349.12) and (871.12,348) .. (872.5,348) .. controls (873.88,348) and (875,349.12) .. (875,350.5) .. controls (875,351.88) and (873.88,353) .. (872.5,353) .. controls (871.12,353) and (870,351.88) .. (870,350.5) -- cycle ;
			\draw  [color={rgb, 255:red, 155; green, 155; blue, 155 }  ,draw opacity=1 ][fill={rgb, 255:red, 155; green, 155; blue, 155 }  ,fill opacity=1 ] (870,327.5) .. controls (870,326.12) and (871.12,325) .. (872.5,325) .. controls (873.88,325) and (875,326.12) .. (875,327.5) .. controls (875,328.88) and (873.88,330) .. (872.5,330) .. controls (871.12,330) and (870,328.88) .. (870,327.5) -- cycle ;
			\draw  [color={rgb, 255:red, 155; green, 155; blue, 155 }  ,draw opacity=1 ] (691,268) -- (726,268) -- (726,303) -- (691,303) -- cycle ;
			\draw  [color={rgb, 255:red, 155; green, 155; blue, 155 }  ,draw opacity=1 ] (693,367) -- (728,367) -- (728,402) -- (693,402) -- cycle ;
			\draw  [color={rgb, 255:red, 184; green, 0; blue, 22 }  ,draw opacity=1 ][line width=2.25]  (567,38) -- (662.5,38) -- (662.5,96) -- (567,96) -- cycle ;
			\draw  [color={rgb, 255:red, 65; green, 117; blue, 5 }  ,draw opacity=1 ][line width=2.25]  (863,39) -- (958.5,39) -- (958.5,97) -- (863,97) -- cycle ;
			\draw  [color={rgb, 255:red, 36; green, 36; blue, 36 }  ,draw opacity=1 ][dash pattern={on 1.69pt off 2.76pt}][line width=1.5]  (557.5,37.4) .. controls (557.5,24.48) and (567.98,14) .. (580.9,14) -- (945.1,14) .. controls (958.02,14) and (968.5,24.48) .. (968.5,37.4) -- (968.5,107.6) .. controls (968.5,120.52) and (958.02,131) .. (945.1,131) -- (580.9,131) .. controls (567.98,131) and (557.5,120.52) .. (557.5,107.6) -- cycle ;
			\draw  [color={rgb, 255:red, 184; green, 0; blue, 22 }  ,draw opacity=1 ][dash pattern={on 6.75pt off 4.5pt}][line width=2.25]  (182,282) -- (254,282) -- (254,340) -- (182,340) -- cycle ;
			\draw  [color={rgb, 255:red, 65; green, 117; blue, 5 }  ,draw opacity=1 ][dash pattern={on 6.75pt off 4.5pt}][line width=2.25]  (1005,226) -- (1093,226) -- (1093,262) -- (1005,262) -- cycle ;
			\draw  [color={rgb, 255:red, 65; green, 117; blue, 5 }  ,draw opacity=1 ][dash pattern={on 6.75pt off 4.5pt}][line width=2.25]  (1005,269) -- (1093,269) -- (1093,305) -- (1005,305) -- cycle ;
			\draw  [color={rgb, 255:red, 65; green, 117; blue, 5 }  ,draw opacity=1 ][dash pattern={on 6.75pt off 4.5pt}][line width=2.25]  (1005,366) -- (1093,366) -- (1093,402) -- (1005,402) -- cycle ;
			\draw  [color={rgb, 255:red, 155; green, 155; blue, 155 }  ,draw opacity=1 ][fill={rgb, 255:red, 155; green, 155; blue, 155 }  ,fill opacity=1 ] (1046,314.5) .. controls (1046,313.12) and (1047.12,312) .. (1048.5,312) .. controls (1049.88,312) and (1051,313.12) .. (1051,314.5) .. controls (1051,315.88) and (1049.88,317) .. (1048.5,317) .. controls (1047.12,317) and (1046,315.88) .. (1046,314.5) -- cycle ;
			\draw  [color={rgb, 255:red, 155; green, 155; blue, 155 }  ,draw opacity=1 ][fill={rgb, 255:red, 155; green, 155; blue, 155 }  ,fill opacity=1 ] (1046,356.5) .. controls (1046,355.12) and (1047.12,354) .. (1048.5,354) .. controls (1049.88,354) and (1051,355.12) .. (1051,356.5) .. controls (1051,357.88) and (1049.88,359) .. (1048.5,359) .. controls (1047.12,359) and (1046,357.88) .. (1046,356.5) -- cycle ;
			\draw  [color={rgb, 255:red, 155; green, 155; blue, 155 }  ,draw opacity=1 ][fill={rgb, 255:red, 155; green, 155; blue, 155 }  ,fill opacity=1 ] (1046,333.5) .. controls (1046,332.12) and (1047.12,331) .. (1048.5,331) .. controls (1049.88,331) and (1051,332.12) .. (1051,333.5) .. controls (1051,334.88) and (1049.88,336) .. (1048.5,336) .. controls (1047.12,336) and (1046,334.88) .. (1046,333.5) -- cycle ;
			\draw [color={rgb, 255:red, 155; green, 155; blue, 155 }  ,draw opacity=1 ][line width=1.5]    (1095,245) -- (1291,245) -- (1342.33,293.93) ;
			\draw [shift={(1344.5,296)}, rotate = 223.63] [fill={rgb, 255:red, 155; green, 155; blue, 155 }  ,fill opacity=1 ][line width=1.5]  [draw opacity=0] (11.61,-5.58) -- (0,0) -- (11.61,5.58) -- cycle    ;
			
			\draw [color={rgb, 255:red, 155; green, 155; blue, 155 }  ,draw opacity=1 ][line width=1.5]    (1093,291) -- (1289,291) -- (1325.11,301.19) ;
			\draw [shift={(1328,302)}, rotate = 195.75] [fill={rgb, 255:red, 155; green, 155; blue, 155 }  ,fill opacity=1 ][line width=1.5]  [draw opacity=0] (11.61,-5.58) -- (0,0) -- (11.61,5.58) -- cycle    ;
			
			\draw [color={rgb, 255:red, 155; green, 155; blue, 155 }  ,draw opacity=1 ][line width=1.5]    (1096,388) -- (1292,388) -- (1342.47,333.21) ;
			\draw [shift={(1344.5,331)}, rotate = 492.65] [fill={rgb, 255:red, 155; green, 155; blue, 155 }  ,fill opacity=1 ][line width=1.5]  [draw opacity=0] (11.61,-5.58) -- (0,0) -- (11.61,5.58) -- cycle    ;
			
			\draw [color={rgb, 255:red, 155; green, 155; blue, 155 }  ,draw opacity=1 ][line width=1.5]    (253.5,312) -- (419,312) ;
			\draw [shift={(422,312)}, rotate = 180] [fill={rgb, 255:red, 155; green, 155; blue, 155 }  ,fill opacity=1 ][line width=1.5]  [draw opacity=0] (11.61,-5.58) -- (0,0) -- (11.61,5.58) -- cycle    ;
			
			\draw  [color={rgb, 255:red, 36; green, 36; blue, 36 }  ,draw opacity=1 ][dash pattern={on 1.69pt off 2.76pt}][line width=1.5]  (173.5,245.2) .. controls (173.5,220.24) and (193.74,200) .. (218.7,200) -- (1319.8,200) .. controls (1344.76,200) and (1365,220.24) .. (1365,245.2) -- (1365,380.8) .. controls (1365,405.76) and (1344.76,426) .. (1319.8,426) -- (218.7,426) .. controls (193.74,426) and (173.5,405.76) .. (173.5,380.8) -- cycle ;
			\draw [color={rgb, 255:red, 155; green, 155; blue, 155 }  ,draw opacity=1 ][line width=1.5]    (664,70) -- (715,70) ;
			\draw [shift={(718,70)}, rotate = 180] [fill={rgb, 255:red, 155; green, 155; blue, 155 }  ,fill opacity=1 ][line width=1.5]  [draw opacity=0] (11.61,-5.58) -- (0,0) -- (11.61,5.58) -- cycle    ;
			
			\draw [color={rgb, 255:red, 155; green, 155; blue, 155 }  ,draw opacity=1 ][line width=1.5]    (813,68) -- (861,68) ;
			\draw [shift={(864,68)}, rotate = 180] [fill={rgb, 255:red, 155; green, 155; blue, 155 }  ,fill opacity=1 ][line width=1.5]  [draw opacity=0] (11.61,-5.58) -- (0,0) -- (11.61,5.58) -- cycle    ;

			\draw (467,44) node [scale=1.7280000000000002]  {$e( t) =a_{0}\sin( \omega t)$};
			\draw (690,46) node [scale=1.7280000000000002]  {$u_r( t) $};
			\draw (1146,41) node [scale=1.7280000000000002]  {$y( t) =\sum\limits ^{\infty }_{n=1} a_{n}( \omega )\sin( n\omega t+\varphi _{n}( \omega ))$};
			\draw (769,114) node [scale=1.3] [align=left] {{\fontfamily{ptm}\selectfont \textbf{Nonlinear Part }}};
			\draw (471,312) node [scale=1.35,color={rgb, 255:red, 74; green, 74; blue, 74 }  ,opacity=1 ] [align=left] { \ {\fontfamily{ptm}\selectfont \textbf{Virtual}}\\{\fontfamily{ptm}\selectfont \textbf{Harmonic}}\\{\fontfamily{ptm}\selectfont \textbf{Generator}}};
			\draw (88,296) node [scale=1.7280000000000002,color={rgb, 255:red, 0; green, 0; blue, 0 }  ,opacity=1 ]  {$e( t) =a_{0}\sin( \omega t)$};
			\draw (593,231) node [scale=1.7280000000000002,color={rgb, 255:red, 74; green, 74; blue, 74 }  ,opacity=1 ]  {$b_{0}\sin( \omega t+\theta _{0})$};
			\draw (600,269) node [scale=1.7280000000000002,color={rgb, 255:red, 74; green, 74; blue, 74 }  ,opacity=1 ]  {$b_{0}\sin( 2\omega t+\theta _{0})$};
			\draw (604,366) node [scale=1.7280000000000002,color={rgb, 255:red, 74; green, 74; blue, 74 }  ,opacity=1 ]  {$b_{0}\sin( n\omega t+\theta _{0})$};
			\draw (708.5,246.5) node [scale=1.5,color={rgb, 255:red, 74; green, 74; blue, 74 }  ,opacity=1 ]  {$\mathbf{H_{1}}$};
			\draw (844,230) node [scale=1.7280000000000002,color={rgb, 255:red, 74; green, 74; blue, 74 }  ,opacity=1 ]  {$b_{1}( \omega )\sin( \omega t+\theta _{0} +\theta _{1})$};
			\draw (858,270) node [scale=1.7280000000000002,color={rgb, 255:red, 74; green, 74; blue, 74 }  ,opacity=1 ]  {$b_{2}( \omega )\sin( 2\omega t+2\theta _{0} +\theta _{2})$};
			\draw (860,372) node [scale=1.7280000000000002,color={rgb, 255:red, 74; green, 74; blue, 74 }  ,opacity=1 ]  {$b_{n}( \omega )\sin( n\omega t+n\theta _{0} +\theta _{n})$};
			\draw (1344,311) node [scale=1.5,color={rgb, 255:red, 155; green, 155; blue, 155 }  ,opacity=1 ]  {$\bm{\sum }$};
			\draw (1540,285) node [scale=1.7280000000000002]  {$y( t) =\sum\limits ^{\infty }_{n=1} a_{n}( \omega )\sin( n\omega t+\varphi _{n}( \omega ))$};
			\draw (708.5,285.5) node [scale=1.5,color={rgb, 255:red, 74; green, 74; blue, 74 }  ,opacity=1 ]  {$\mathbf{H_{2}}$};
			\draw (710.5,384.5) node [scale=1.5,color={rgb, 255:red, 74; green, 74; blue, 74 }  ,opacity=1 ]  {$\mathbf{H_{n}}$};
			\draw (765.75,67) node [scale=3]  {$\mathfrak{C}_{\mathfrak{R}}$};
			\draw (618,114) node [scale=1.3] [align=left] {{\fontfamily{ptm}\selectfont \textbf{Linear Part }}};
			\draw (614.75,67) node [scale=3]  {$\mathfrak{C}_{L_{1}}$};
			\draw (914,114) node [scale=1.3] [align=left] {{\fontfamily{ptm}\selectfont \textbf{Linear Part }}};
			\draw (910.75,67) node [scale=3]  {$\mathfrak{C}_{L_{2}}$};
			\draw (218.75,312) node [scale=1.5,color={rgb, 255:red, 74; green, 74; blue, 74 }  ,opacity=1 ]  {$C_{L\mathbf{_{1}}}( j\omega )$};
			\draw (1050,244) node [scale=1.7,color={rgb, 255:red, 74; green, 74; blue, 74 }  ,opacity=1 ]  {$C_{L\mathbf{_{2}}}( j\omega )$};
			\draw (1050,287) node [scale=1.7,color={rgb, 255:red, 74; green, 74; blue, 74 }  ,opacity=1 ]  {$C_L{\mathbf{_{2}}}( 2j\omega )$};
			\draw (1050,384) node [scale=1.7,color={rgb, 255:red, 74; green, 74; blue, 74 }  ,opacity=1 ]  {$C_L{\mathbf{_{2}}}( nj\omega )$};
			\draw (1190,229) node [scale=1.7280000000000002,color={rgb, 255:red, 74; green, 74; blue, 74 }  ,opacity=1 ]  {$a_{1}( \omega )\sin( \omega t+\varphi _{1})$};
			\draw (1195,275) node [scale=1.7280000000000002,color={rgb, 255:red, 74; green, 74; blue, 74 }  ,opacity=1 ]  {$a_{2}( \omega )\sin( 2\omega t+\varphi _{2})$};
			\draw (1197,373) node [scale=1.7280000000000002,color={rgb, 255:red, 74; green, 74; blue, 74 }  ,opacity=1 ]  {$a_{n}( \omega )\sin( n\omega t+\varphi_n)$};
			\draw (337,293) node [scale=1.7280000000000002,color={rgb, 255:red, 74; green, 74; blue, 74 }  ,opacity=1 ]  {$b_{0}\sin( \omega t+\theta _{0})$};
			\end{tikzpicture}
		}
		\caption{HOSIDF Representation}
		\label{F-01}
	\end{figure*}
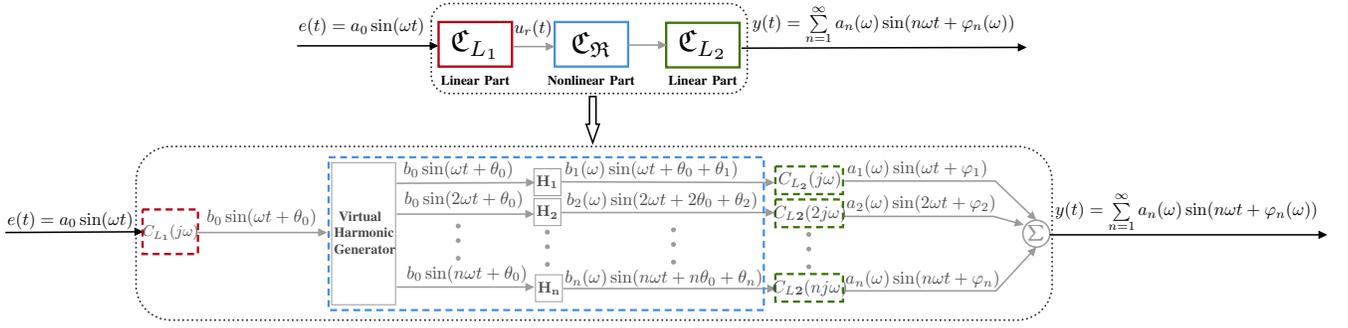
	
	The HOSIDF tool shows that the plant, as well as the linear parts of the controller, influence the high order harmonics. Further, although traditionally the nonlinear reset element is placed to receive error signal as its input, changing the sequence of linear parts and nonlinear reset elements results in different high order harmonic shapes which should result in different resetting laws and closed-loop performance. However, the effects of this sequence on the performance of reset systems have not been investigated so far. In this paper, the effects of different sequences of controller parts on the performance of reset systems are studied using the HOSIDF tool. The best sequence is selected from HOSIDF theory to achieve the highest precision while also ensuring the lowest magnitude control input. This sequence is then tested for closed-loop performance in simulation and on a high precision positioning setup.  
	
	In section \ref{two}, relevant preliminaries of reset control and frequency domain tools are presented. Theoretical investigation of different sequences of controller parts is presented in section \ref{three}. In section \ref{four}, the simulation results from closed-loop for the different sequences are analysed. The experimental results and conclusion are described in sections \ref{five} and \ref{six}, separately.
	
	\section{PRELIMINARIES}\label{two}
	
	\subsection{Reset Control}
	A general SISO reset element is defined by the following state-space equations according to \cite{banos2011reset} as:
	\begin{equation}
	\Sigma_R= \begin{cases}
	\dot{x}_r(t)={A_r}{x_r}(t)+{B_r}u_r(t)  & \mbox{if}\ u_r(t)\ne0  \\
	x_{r}(t^+)=A_{\rho}{x_r}(t)   & \mbox{if}\ u_r(t)=0 \\
	y_r(t)={C_r}{x_r}(t)+{D_r}u_r(t)
	\end{cases}
	\label{general reset controller}
	\end{equation}
	where $A_r$, $B_r$, $C_r$ and $D_r$ are state-space matrices of the corresponding base linear system, $A_\rho$ is the reset matrix determining the states' value ($x_r(t^+)$) after reset action, $u_r(t)$ is the input and is traditionally the error signal $e(t)$ and $y_r(t)$ is the output. To simplify the design of the reset element, reset matrix $A_\rho$ is often defined as a diagonal matrix:
	\begin{equation}\label{E2}
	A_\rho=\gamma I_{n_r\times n_r},\quad \gamma\in[-1,1]
	\end{equation}
	where $n_r$ is the order of the reset controller. Although several reset laws exist in literature, we utilize the zero input crossing, i.e., $u_r(t) = 0$ as the reset law in this paper. To avoid Zeno behaviour, two consecutive reset instants are prevented.  
	\newline DF of the defined reset element for a sinusoidal input is obtained in \cite{guo2009frequency} as:
	\begin{equation}
	\mathcal{N}_{DF}=C_r^T(j\omega I-A_r)^{-1}(I+j\Theta_\rho(\omega))B_r+D_r
	\label{DF}
	\end{equation}
	where
	\begin{equation}
	\Theta_\rho=\dfrac{2}{\pi}\dfrac{(I+e^{\frac{\pi A_r}{\omega}})}{(I+A_\rho e^\frac{\pi A_r}{\omega})}\dfrac{(I-A_\rho)}{((\frac{A_r}{\omega})^2+I)}
	\end{equation}
	
	In addition, HOSIDF for general reset elements are obtained in \cite{heinen2018frequency} as: 
	\begin{equation}
	\resizebox{\hsize}{!}{$
		H_n(j\omega)=\begin{cases}
		C_r(j\omega I-A_r)^{-1}(I+j\Theta_\rho(\omega))B_r+D_r  & \ n=1  \\
		C_r(j\omega nI-A_r)^{-1}j\Theta_\rho(\omega)B_r   & \text{odd}\ n\ge2 \\
		0   &  \text{even}\ n\ge2
		\end{cases}$}
	\label{HOSIDF}
	\end{equation}
	where $n$ is the order of the harmonic.
	
	The linear part of the controller which receives the error input is defined as $C_{L_1}$ and the linear part following the reset element is defined as $C_{L_2}$. This is as shown in Fig. \ref{F-01}. If the input of reset element is error ($C_{L_1}=1$), then it results in the zero error crossing as introduced by Clegg. If the plant is defined as $G$, then the DF and HOSIDF of the open-loop $L$ is obtained as
	\begin{equation}
	L_n(j\omega)= C_{L_1}(j\omega)H_n(j\omega)C_{L_2}(nj\omega)G(nj\omega)
	\label{DFHOSIDFol}
	\end{equation}
	
	\subsection{Pseudo Sensitivity Functions}
	In linear systems, tracking error is obtained through sensitivity function which is defined as:
	\begin{equation}
	\frac { e } { r } = S ( j \omega ) = \frac { 1 } { 1 + L( j \omega ) }
	\label{old sensitivity}
	\end{equation}
	where $L ( j \omega )$ is the open loop frequency response of the linear system.
	
	In order to get sensitivity function of nonlinear systems, DF can be used to get $L_1( j \omega )$. However, DF only considers the first harmonic. To take into account the influence of high order harmonics, from a precision perspective, a pseudo-sensitivity function is defined for nonlinear systems as the ratio of the maximum tracking error of the system to the magnitude of the reference at each frequency. In other words,
	\begin{equation}
	\forall \omega\in\mathbb{R}^{+}:\ S _ { \partial } ( \omega ) = {\frac { \max ( | e ( t ) | ) } { | r | }} \qquad  \mbox{for}\  {t \geq t _ { ss }}
	\label{new sensitivity}
	\end{equation}
	where $t _ { ss }$ is the time when system reaches steady state and $r$ is the amplitude of sinusoidal reference input. Since $\max ( | e ( t ) | )$ is the summation error of all the harmonics, this pseudo sensitivity function is more reliable than \eqref{old sensitivity} for nonlinear controllers and will be used for closed-loop performance analyses.
	
	\section{METHODOLOGY}\label{three}
	In linear controllers, the sequence of the different linear filters does not affect the performance since they result in the same transfer function. However, when reset elements are used, the performance of the system can vary depending on the relative sequence of controller parts because the magnitude of high order harmonics depends on the chosen sequence and this in-turn influences the closed-loop performance. As shown in (\ref{DFHOSIDFol}), while the DF (when $n = 1$) is not affected by the sequence, the magnitude of high order harmonics of the whole controller are influenced by $\mathfrak{C}_{L_1}$, $\mathfrak{C}_{L_2}$ and even the plant $G$. Therefore, HOSIDF tool is used to investigate and compare the magnitude of high order harmonics of different sequences.
	
	
	In general, linear controllers can be divided into lead $\mathfrak {C}_{lead}$ and lag $\mathfrak {C}_{lag}$ filters. With the inclusion of the reset element, resulting in three controller parts, there are six different sequences possible. However, if linear lead and lag elements are interchanged, no difference will be seen in performance. Hence the number of sequences for investigation reduces to four and these are listed in TABLE~\ref{T1}.  
	
	\begin{table}[h]
		\caption{Different sequences of general case}
		\centering
		\begin{tabular}{|c||c||c|}
			\hline
			\textbf{No.} & \textbf{Sequence} & \textbf{nth order harmonic}\\
			\hline  
			1  & \color{rgb, 255:red, 65; green, 117; blue, 5 }{\textbf{Lead-Reset-Lag}} & $\mathfrak { C } _ { lead }(j\omega) H_n(j\omega) \mathfrak { C } _ { lag }(nj\omega) G(nj\omega)$\\
			\hline  
			2  & Lag-Reset-Lead & $\mathfrak { C } _ { lag }(j\omega) H_n(j\omega) \mathfrak { C } _ { lead }(nj\omega)  G(nj\omega)$\\
			\hline  
			3  & Reset-Lead-Lag & $H_n(j\omega) \mathfrak { C } _ { lead }(nj\omega) \mathfrak { C } _ { lag }(nj\omega)  G(nj\omega)$\\
			\hline  
			4  & Lead-Lag-Reset & $\mathfrak { C } _ { lead }(j\omega) \mathfrak { C } _ { lag }(j\omega)\times H_n(j\omega) G(nj\omega)$\\
			
			\hline 
		\end{tabular}
		\label{T1}
	\end{table}
	
	Based on the equations in TABLE \ref{T1}, the first harmonic $(n=1)$ or DF for all 4 sequences are the same. However, for high order harmonics, since lead filters are ascending functions in magnitude while lag filters are descending functions with respect to the frequency, it is obvious that the first (No.1) and second (No.2) sequence has the smallest and largest magnitude of high order harmonics, respectively.
	
	\begin{figure}[htbp]
		\centering{\includegraphics[width=\columnwidth]{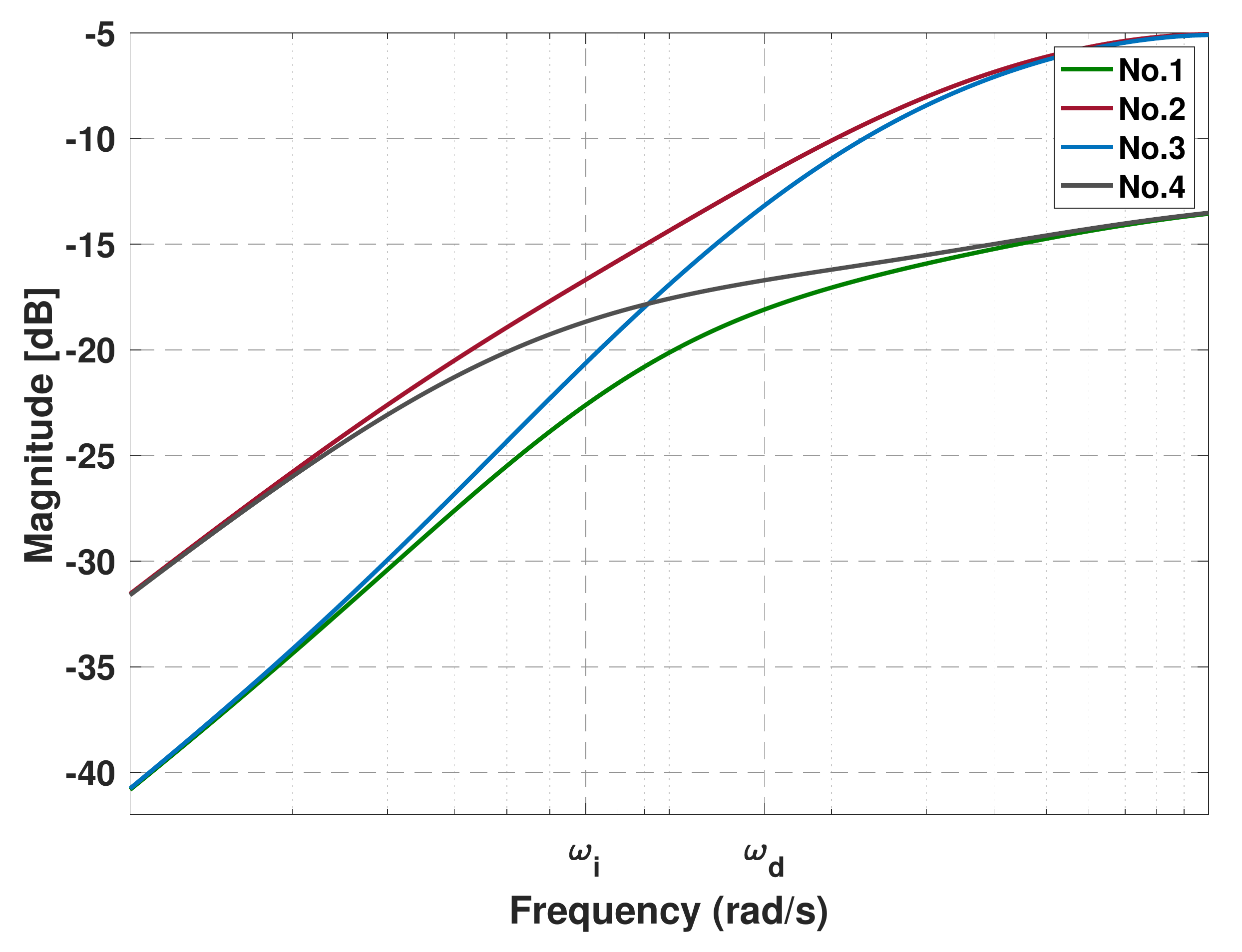}}
		\caption{The magnitude of the third order harmonic for different sequences of FORE}
		\label{general_best_sequence}
	\end{figure}
	
	For a simple example, let us combine a FORE with a first-order lead filter $(1+\frac{s}{\omega_{d}})$ and a first-order lag filter: $(1+\frac{\omega_{i}}{s})$, with the magnitude of the third order harmonic for the four different sequences visualized in Fig.~\ref{general_best_sequence}. It is clear that before $\omega_i$, the lag filter plays a role so that $No. 1$ and $No. 3$ have a smaller magnitude of high order harmonics. After $\omega_i$, the lag filter effect has been terminated and the lead filter comes into play, therefore, $No. 1$ and $No. 4$ become smaller.  In all range of frequencies, $No.1$ always has the smallest magnitude while $No. 2$ has the largest magnitude of high order harmonics. The other two sequences are a trade-off between the extremes. Since the high order harmonics deteriorate the closed-loop performance, the optimal sequence is hypothesized to be the one with the lowest magnitude of high order harmonics. Based on HOSIDF theory, we can say that the optimal sequence for reset systems results in all linear lead elements preceding and all linear lag elements following the reset element i.e., $No. 1$.
	
	\section{CLOSED-LOOP PERFORMANCE}\label{four}
	
	To validate our hypothesis and investigate the closed-loop performances of different sequences, a Lorentz-actuated precision positioning stage is used as a benchmark.
	
	\subsection{System Overview}
\begin{figure}[htbp]
		\centering{\includegraphics[width=0.6\columnwidth]{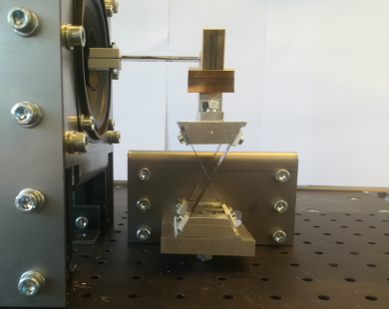}}
		\caption{Precision positioning stage actuated by a loud speaker}
		\label{setup}
	\end{figure}

\begin{figure}[htbp]
		\centering{\includegraphics[width=\columnwidth]{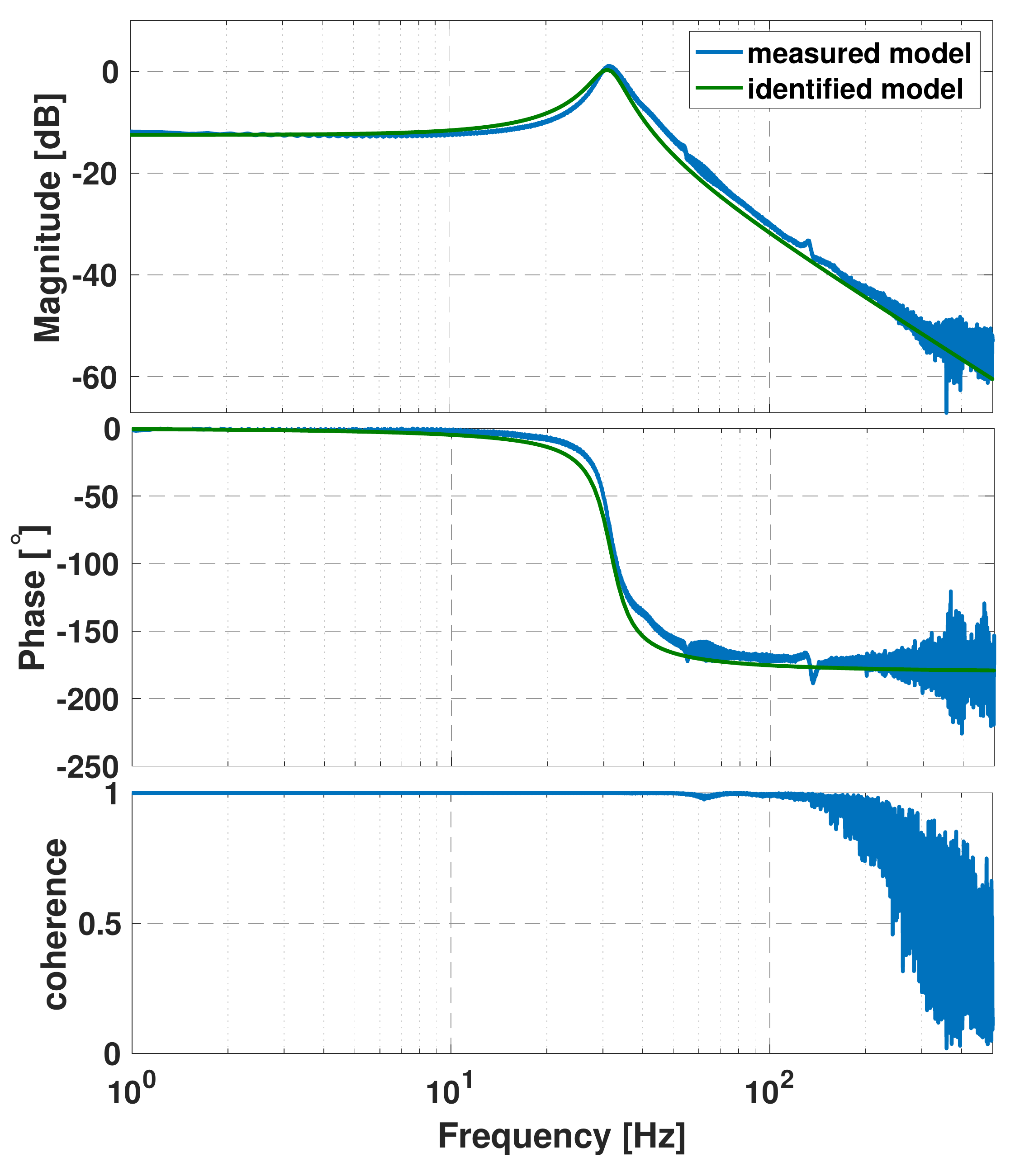}}
		\caption{Frequency response of the system identification}
		\label{identify}
	\end{figure}

	The system shown in Fig.~\ref{setup} consists of a mass guided using flexure cross hinge and actuated by a Visaton FR10-4 loudspeaker. With a Mercury 2000 reflective linear encoder, the horizontal position of the stage is measured with a resolution of 100nm. The controllers are implemented using FPGA module via compact RIO real-time hardware. Fig.~\ref{identify} shows the frequency response of the system. This system is identified as a second order mass-spring-damper system with the transfer function:
	\begin{equation}
	P(s)=\frac{1}{1.077\times10^{-4}s^2+0.0049s+4.2218}.
	\end{equation}
	
	\subsection{Controller Design}
\begin{figure}[htbp]
		\centering{\includegraphics[width=.85\columnwidth]{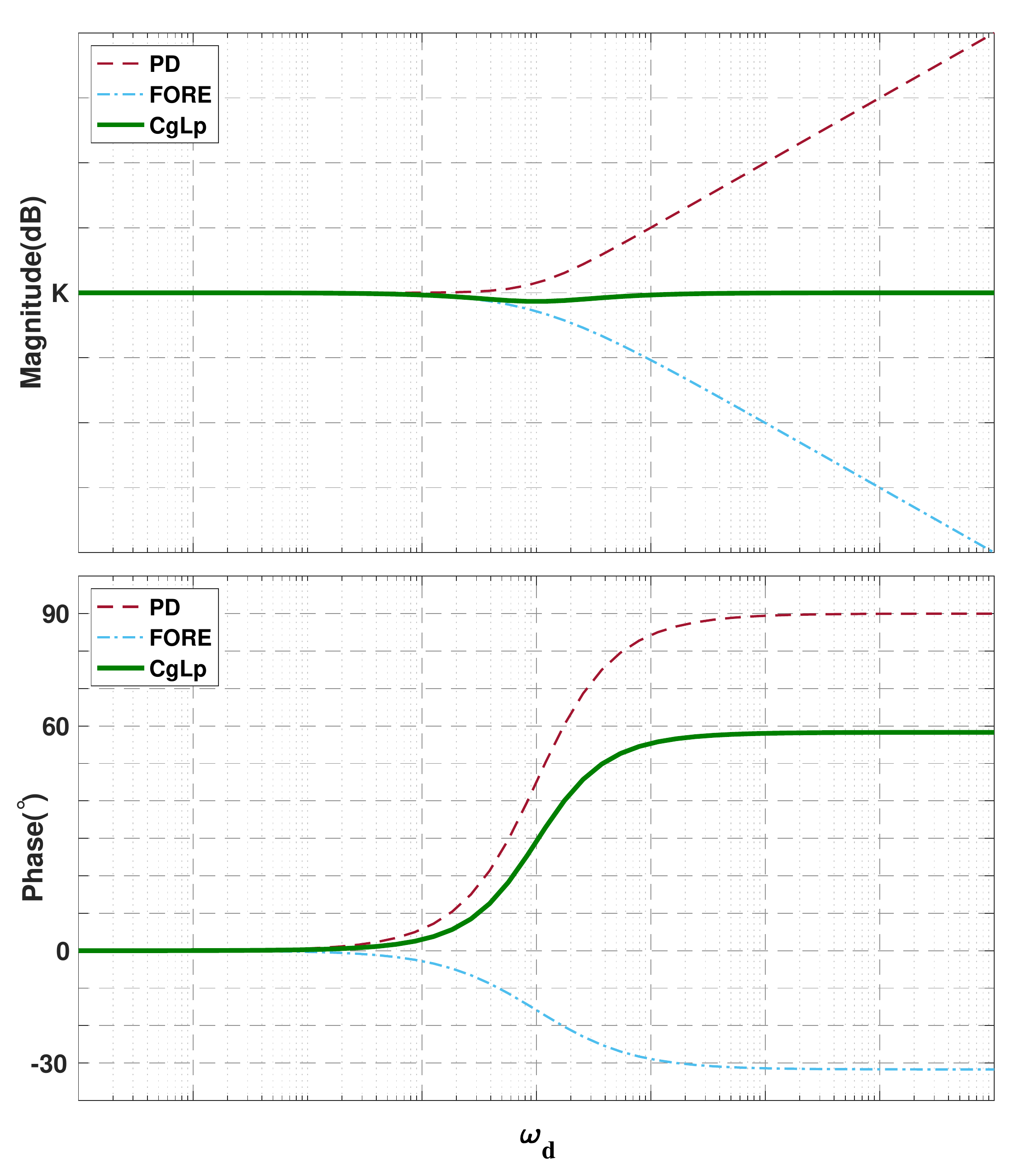}}
		\caption{The DF based frequency behavior a (CgLp) compensator}
		\label{Bode_CgLp}
	\end{figure}

	For controlling this system, a Proportional Integration (PI) with a Constant in gain Lead in phase (CgLp) compensator used. CgLp element is made up of a reset lag filter and a corresponding linear lead filter as proposed in \cite{saikumar2019constant}. Consider a FORE and a linear lead part $D$ as given below:
	\begin{equation}
	\text{FORE}(s)=\frac{1}{\cancelto{\gamma}{s/\omega_{r}+1}}
	\end{equation}
	and
	\begin{equation}
	D(s)=\frac { s / \omega _ { d } + 1 } { s / \omega _ { t } + 1 }
	\end{equation}
	where $\omega_{r}$ is the corner frequency of reset element, $\gamma$ determines the reset value (as defined in (\ref{E2})), $\omega_d$ and $\omega_t$ are starting and taming frequencies of linear lead filter. By tuning $\omega_{r}=\omega_d/\alpha$, where $\alpha$ is a correction factor chosen according to \cite{saikumar2019constant}, broadband phase lead can be achieved in the range of $[\omega_d,\omega_t]$ with constant gain (based on DF) as shown in Fig.~\ref{Bode_CgLp}.
	By replacing the $D$ part of a traditional PID controller with a CgLp element, PI+CgLp controller is defined as:
	\begin{equation}
	\Sigma _ { R C }= \underbrace { K _ { p } \left( 1 + \frac { \omega _ { i } } { s } \right)} _ { PI } \underbrace {\left( \frac { 1 + \frac { s } { \omega _ { d } } } { 1 + \frac { s } { \omega _ { t } } } \right) \left(\frac{1}{\cancelto{\gamma}{s/\omega_{r}+1}}\right)} _ { CgLp }
	\end{equation}
	where $\omega_i$ is the corner frequency of the integrator element, and $K_p$ is the proportional gain.
	
	Based on the DF, controllers are designed to have the cross-over frequency $\omega_c=100$Hz with $30^{\circ}$ phase margin. $\gamma$ is selected as zero (classical reset), and according to \cite{Hou,saikumar2019constant}, $\omega_d$ is chosen to be $\omega_c/4$, $\omega_i$=$\omega_c/10$ and $\omega_t$=$6\omega_c$, and correction factor $\alpha$ is taken as 1.62 ($\omega_{r}=\omega_d/1.62$). Also, $K_p$ is tuned to get the required cross-over frequency. The parameters of the controller are listed in TABLE~\ref{Parameters of controller}.
	\begin{table}[h]
		\caption{Tuning parameters of PI+CgLp controller}
		\centering
		\begin{tabular}{|c||c||c|}
			\hline
			\textbf{symbol} & \textbf{parameter} & \textbf{Value}\\
			\hline  
			$\omega_c$  & bandwidth & 100 Hz\\
			\hline 
			$\omega_d$  & corner frequency of lead filter & 25Hz\\
			\hline 
			$\omega_t$  & taming frequency of lead filter & 600 Hz\\
			\hline 
			$\omega_{r}$ &corner frequency of reset lag filter & 15.43 Hz\\
			\hline 
			$\omega_i$ & corner frequency of integrator & 10 Hz\\
			\hline 
			$K_p$ & proportional gain of the controller & 3980 \\
			\hline 
			
		\end{tabular}
		\label{Parameters of controller}
	\end{table}
	A PI+CgLp consists of a lag element (PI), a lead element (D) and a reset element (FORE). As TABLE~\ref{T1}, four relative sequences are to be considered.
	\subsection{Closed-Loop Performance Analysis in simulation}
	
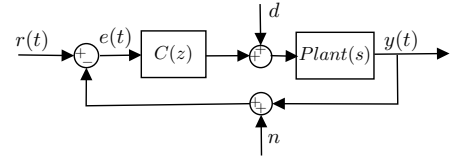
\begin{figure}[htbp]
		\centering
		\resizebox{.7\hsize}{!}{
			\tikzset{every picture/.style={line width=0.75pt}} 
			\begin{tikzpicture}[x=0.75pt,y=0.75pt,yscale=-1,xscale=1]
			
			\draw   (166,58) -- (211.67,58) -- (211.67,91) -- (166,91) -- cycle ;
			\draw   (275.54,59) -- (330.54,59) -- (330.54,92) -- (275.54,92) -- cycle ;
			\draw    (211.67,75.67) -- (240.88,75.51) ;
			\draw [shift={(242.88,75.5)}, rotate = 539.69] [fill={rgb, 255:red, 0; green, 0; blue, 0 }  ][line width=0.75]  [draw opacity=0] (8.93,-4.29) -- (0,0) -- (8.93,4.29) -- cycle    ;
			
			\draw    (257.41,75.5) -- (275,75.95) ;
			\draw [shift={(277,76)}, rotate = 181.46] [fill={rgb, 255:red, 0; green, 0; blue, 0 }  ][line width=0.75]  [draw opacity=0] (8.93,-4.29) -- (0,0) -- (8.93,4.29) -- cycle    ;
			
			\draw    (330.83,74.5) -- (382,74.02) ;
			\draw [shift={(384,74)}, rotate = 539.46] [fill={rgb, 255:red, 0; green, 0; blue, 0 }  ][line width=0.75]  [draw opacity=0] (8.93,-4.29) -- (0,0) -- (8.93,4.29) -- cycle    ;
			
			\draw    (347,75) -- (347.36,109.74) -- (260.59,109.51) ;
			\draw [shift={(258.59,109.5)}, rotate = 360.15999999999997] [fill={rgb, 255:red, 0; green, 0; blue, 0 }  ][line width=0.75]  [draw opacity=0] (8.93,-4.29) -- (0,0) -- (8.93,4.29) -- cycle    ;
			
			\draw    (127.15,86) -- (127,111) -- (242.88,109.5) ;
			
			\draw [shift={(127.17,84)}, rotate = 90.35] [fill={rgb, 255:red, 0; green, 0; blue, 0 }  ][line width=0.75]  [draw opacity=0] (8.93,-4.29) -- (0,0) -- (8.93,4.29) -- cycle    ;
			\draw   (118.56,75.5) .. controls (118.56,70.81) and (122.41,67) .. (127.17,67) .. controls (131.92,67) and (135.78,70.81) .. (135.78,75.5) .. controls (135.78,80.19) and (131.92,84) .. (127.17,84) .. controls (122.41,84) and (118.56,80.19) .. (118.56,75.5) -- cycle ;
			\draw    (136.78,75.5) -- (164.26,75.1) ;
			\draw [shift={(166.26,75.07)}, rotate = 539.1700000000001] [fill={rgb, 255:red, 0; green, 0; blue, 0 }  ][line width=0.75]  [draw opacity=0] (8.93,-4.29) -- (0,0) -- (8.93,4.29) -- cycle    ;
			
			\draw   (242.59,76.15) .. controls (242.59,71.81) and (246.1,68.3) .. (250.44,68.3) .. controls (254.78,68.3) and (258.29,71.81) .. (258.29,76.15) .. controls (258.29,80.48) and (254.78,84) .. (250.44,84) .. controls (246.1,84) and (242.59,80.48) .. (242.59,76.15) -- cycle ;
			\draw   (242.88,109.5) .. controls (242.88,105.16) and (246.4,101.65) .. (250.73,101.65) .. controls (255.07,101.65) and (258.59,105.16) .. (258.59,109.5) .. controls (258.59,113.84) and (255.07,117.35) .. (250.73,117.35) .. controls (246.4,117.35) and (242.88,113.84) .. (242.88,109.5) -- cycle ;
			\draw    (251,146.2) -- (250.75,119.35) ;
			\draw [shift={(250.73,117.35)}, rotate = 449.47] [fill={rgb, 255:red, 0; green, 0; blue, 0 }  ][line width=0.75]  [draw opacity=0] (8.93,-4.29) -- (0,0) -- (8.93,4.29) -- cycle    ;
			
			\draw    (250,39) -- (250.41,66.3) ;
			\draw [shift={(250.44,68.3)}, rotate = 269.14] [fill={rgb, 255:red, 0; green, 0; blue, 0 }  ][line width=0.75]  [draw opacity=0] (8.93,-4.29) -- (0,0) -- (8.93,4.29) -- cycle    ;
			
			\draw    (79,75) -- (116.56,75.47) ;
			\draw [shift={(118.56,75.5)}, rotate = 180.72] [fill={rgb, 255:red, 0; green, 0; blue, 0 }  ][line width=0.75]  [draw opacity=0] (8.93,-4.29) -- (0,0) -- (8.93,4.29) -- cycle    ;

			\draw (350,65) node [scale=1.1]  {$y(t)$};
			\draw (188.83,74.5) node [scale=1]  {$C( z)$};
			\draw (303.04,75.5) node [scale=1]  {$Plant( s)$};
			\draw (250.44,71.3) node [scale=0.8]  {$+$};
			\draw (246.44,76.15) node [scale=0.8]  {$+$};
			\draw (250.71,113) node [scale=0.8]  {$+$};
			\draw (246.73,108.5) node [scale=0.8]  {$+$};
			\draw (128.17,80) node [scale=0.8]  {$-$};
			\draw (123.17,73.5) node [scale=0.8]  {$+$};
			\draw (260,134) node [scale=1.1]  {$n$};
			\draw (260,43) node [scale=1.1]  {$d$};
			\draw (89,65) node [scale=1.1]  {$r( t)$};
			\draw (148,63) node [scale=1.1]  {$e( t)$};
			\end{tikzpicture}}
		\caption{Block diagram of closed-loop performance analyses}
		\label{Block diagram}
	\end{figure}

	The defined pseudo-sensitivity function of \eqref{new sensitivity} is used to compare the closed-loop tracking performance in simulation. Disturbance and white noise are added to mimic a more realistic situation as shown in Fig.~\ref{Block diagram}. Control elements are discretized for a sampling frequency of 20 KHz. A disturbance signal between $0.5Hz$ and $30Hz$ which can cause $10\%$ positioning deviation is applied for all sequences to mimic floor vibration. White noise with the magnitude of $(1\%-3\%)$ of the reference is considered to imitate the noise present in the real setup. However, to consider the effect of noise on overall performance, different levels of noise are used during simulation for analysis.
	
\begin{figure*}[htbp]
		\centering
		\subcaptionbox{Sensitivity function with 0.1\% noise \label{f1} }[.36\linewidth]
		{%
			\includegraphics[width=0.36\linewidth]{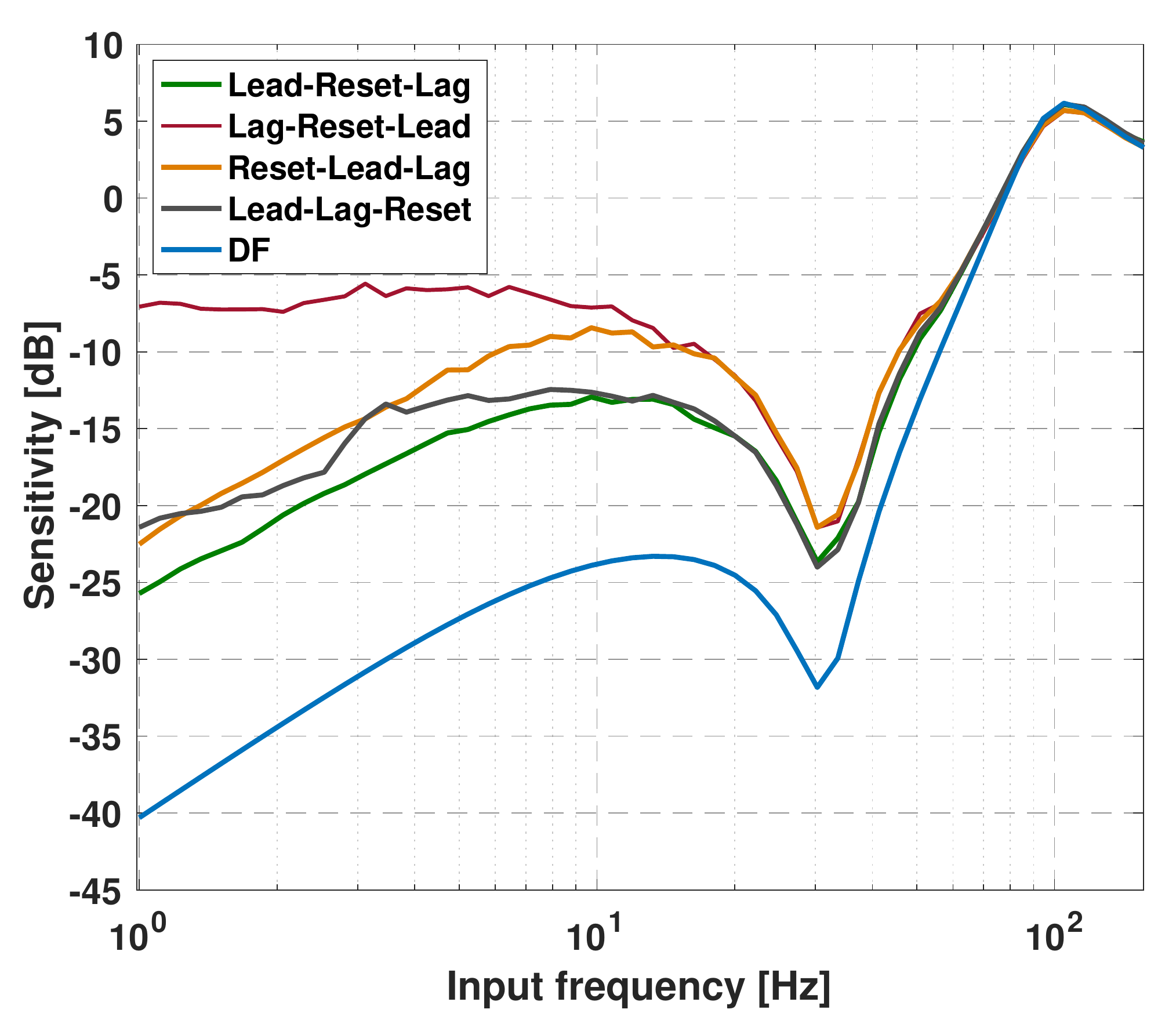} }\quad
		\subcaptionbox{Control output with 0.1\% noise \label{f5}}[.36\linewidth]
		{%
			\includegraphics[width=0.36\linewidth]{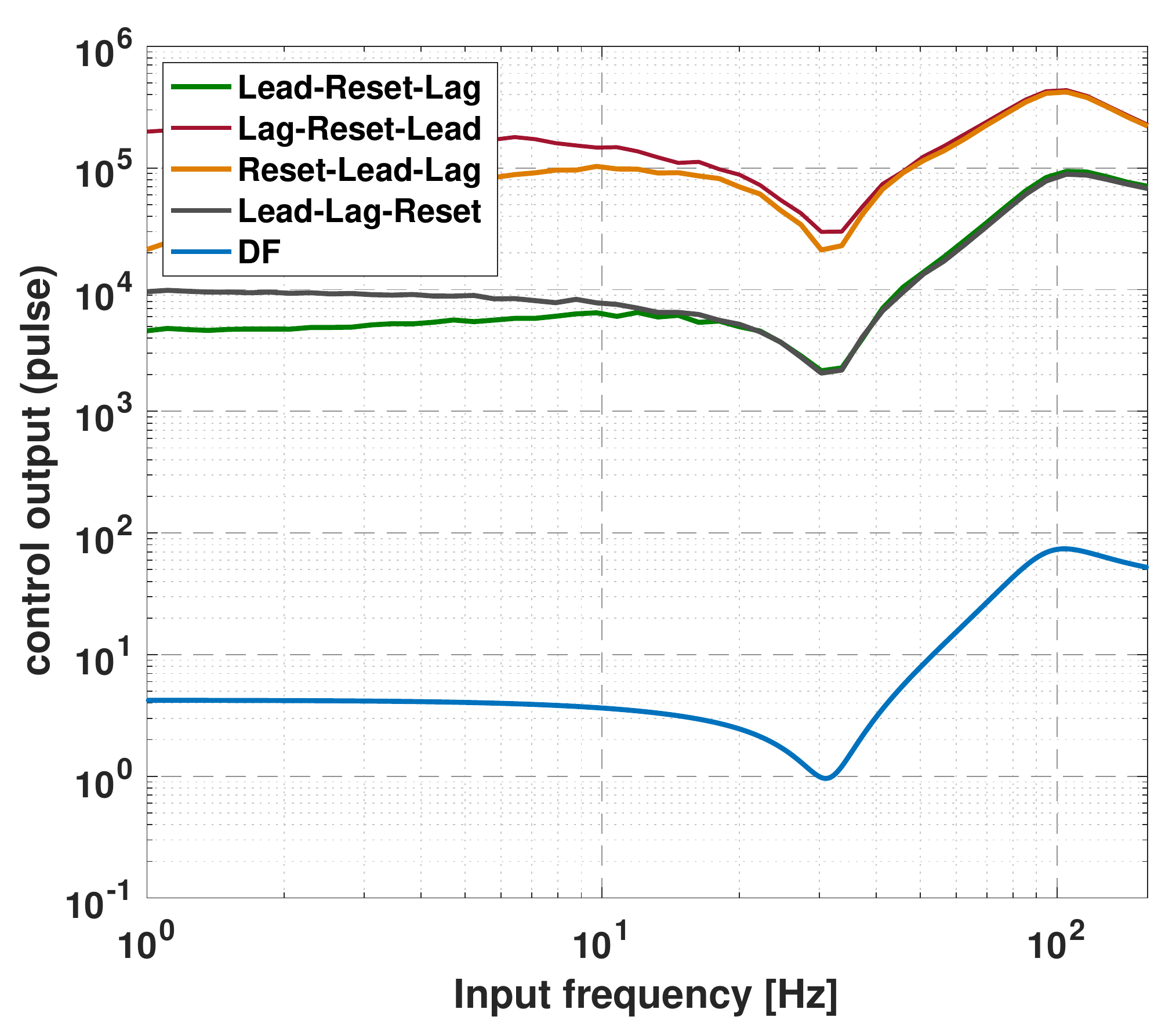} }\\
		
		\subcaptionbox{Sensitivity function with 1\% noise }[.36\linewidth]
		{%
			\includegraphics[width=0.36\linewidth]{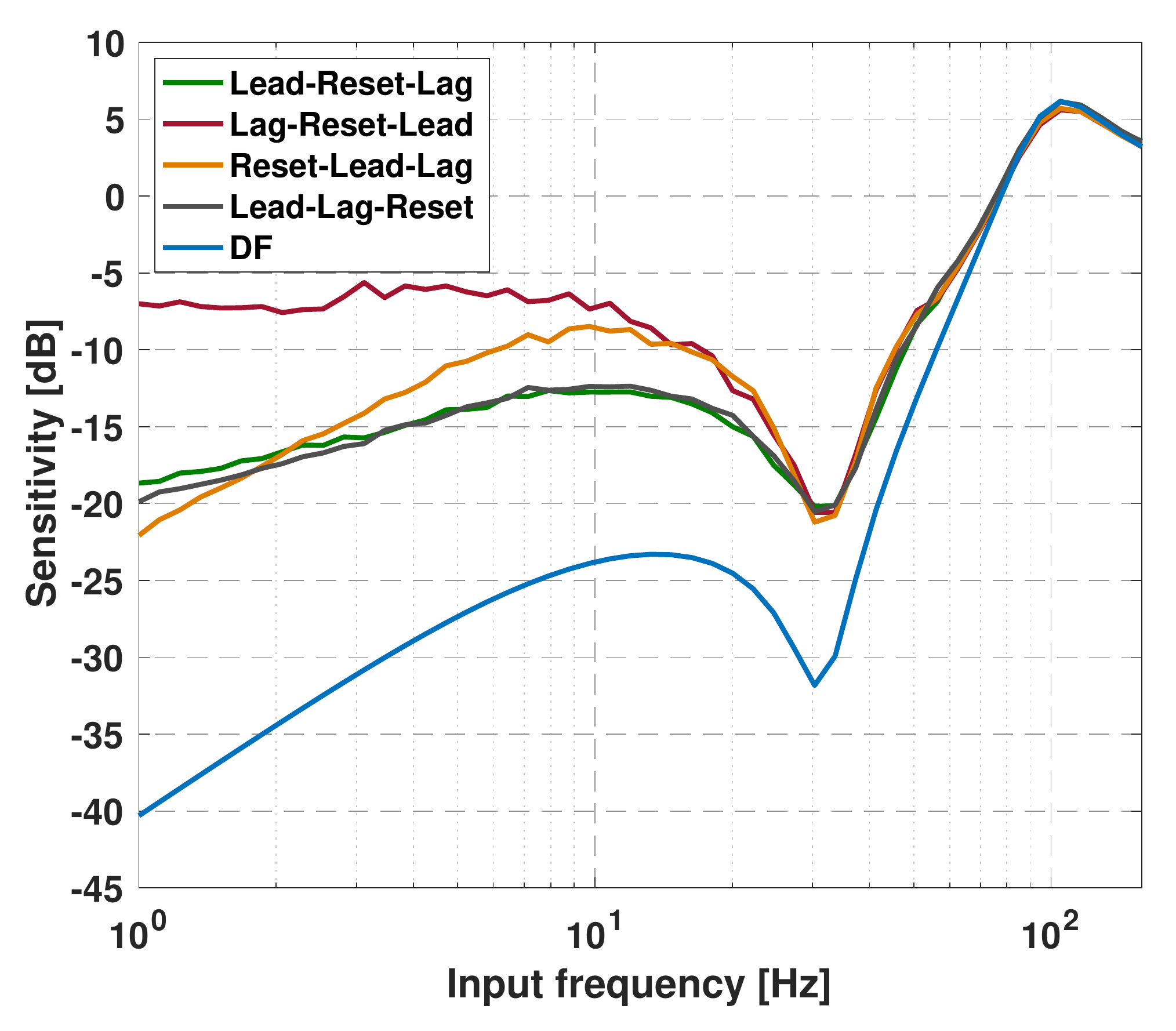} }\quad
		\subcaptionbox{Control output with 1\% noise \label{f2}}[.36\linewidth]
		{%
			\includegraphics[width=0.36\linewidth]{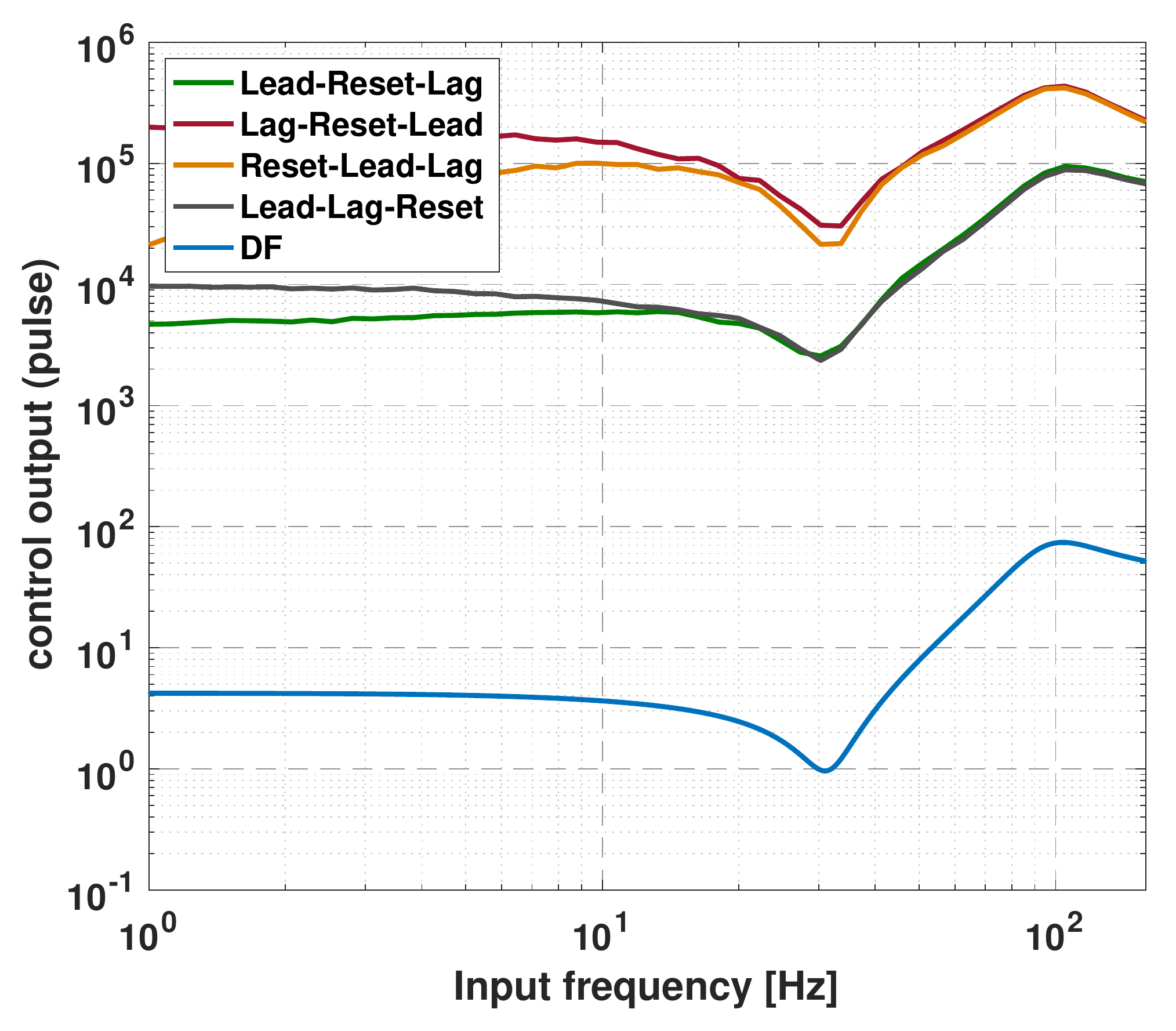} }\\
		
		\subcaptionbox{Sensitivity function with 3\% noise \label{f4}}[.36\linewidth]
		{%
			\includegraphics[width=0.36\linewidth]{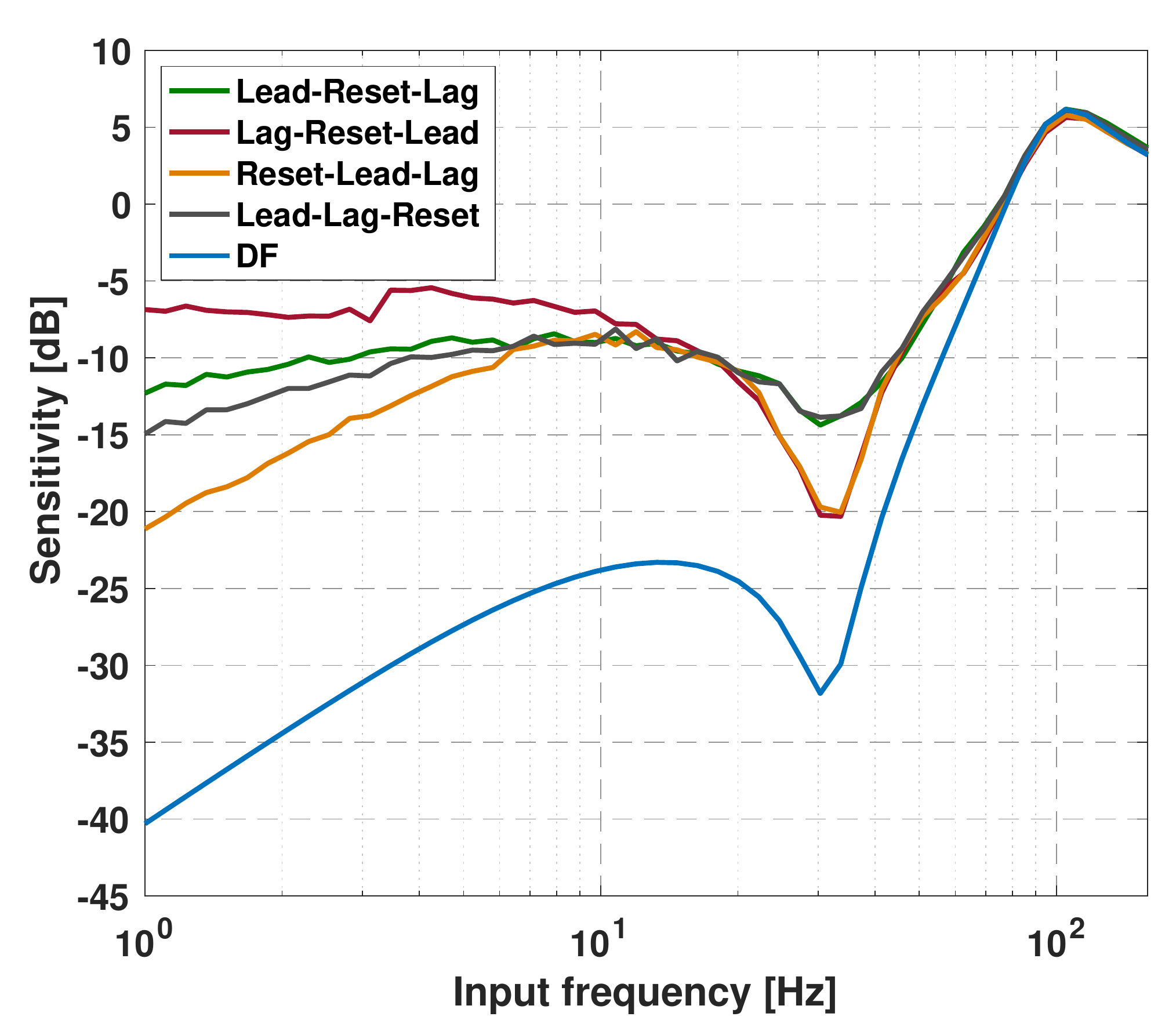} }\quad
		\subcaptionbox{Control output with 3\% noise \label{f8}}[.36\linewidth]
		{%
			\includegraphics[width=0.36\linewidth]{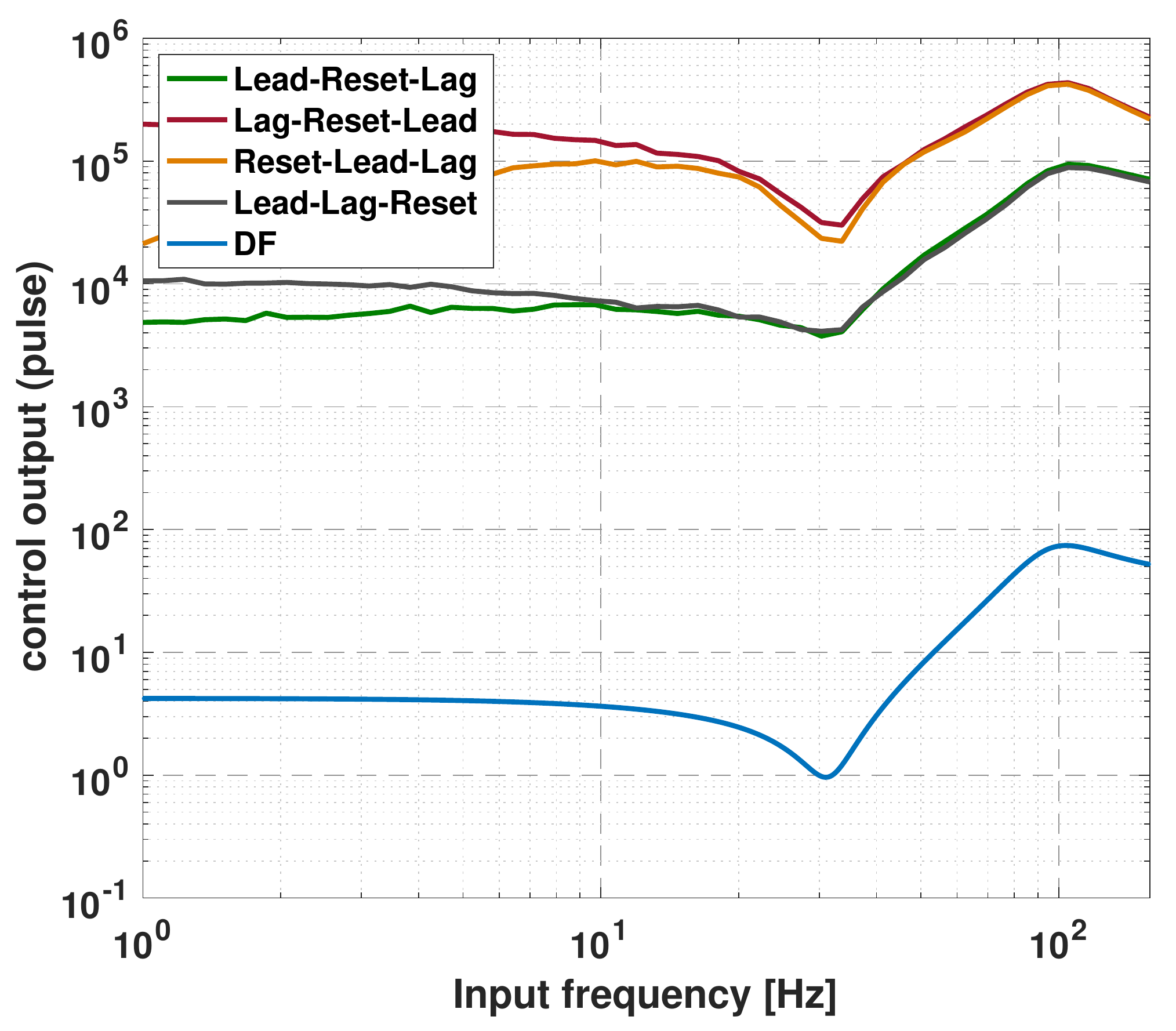} }
		\caption{pseudo-sensitivity functions and maximum control inputs with different level of noise}
		\label{S and C}
	\end{figure*}

	Sinusoidal signal is given as input at different frequencies and the maximum steady-state value of $| e ( t ) |$ was recorded and used to plot $S _ { \partial } ( \omega )$ with different levels of noise as shown in Fig.~\ref{S and C}. Also, the DF sensitivity is plotted using DF of reset element and linear sensitivity relation of (\ref{old sensitivity}). 
	
	The first thing that should be noted is that the DF based sensitivity is not appropriate at estimating closed-loop performance of any of the sequences. Next, concerning the different sequences, as shown in Fig.~\ref{f1}, the sequence  Lead-Reset-Lag ($No. 1$) has the smallest $S _ { \partial } ( \omega )$ at all frequencies when the magnitude of noise signal is $0.1\%$ of reference. When the magnitude of noise increases to more than $1\%$, the performances of sequence $No. 1$ and $No. 4$ deteriorate at low frequencies while the sequences $No. 2$ and $No. 3$ do not change a lot. This deterioration in performance with the increase in noise amplitude can be explained by the fact that both $No. 1$ and $No. 4$ have a lead filter before the reset element which amplifies noise which is present at high frequencies. This amplified noise influences the zero crossing instants. This is especially true at low frequencies of the reference where the error signal in the absence of noise and disturbances would be quite low. In the presence of amplified noise, noise starts dominating the combined error signal. As a result, the zero crossings are dominantly determined by noise resulting in the performance deterioration seen. At higher frequencies especially above $\omega_d$ (25Hz), the error due to reference is also amplified by the lead filter hence cancelling out the detrimental influence of noise.
	
	In terms of the control input, since the maximum amplitude is important for avoiding saturation, the maximum control input values at each frequency are compared for all sequences. As shown in Fig.~\ref{f2}, Fig. \ref{f5}, and Fig. \ref{f8}, sequences $No. 2$ and $No. 3$ always have much larger control input compared with others. This is because these two structures have a lead filter after the reset element. In these sequences, the resetting action which results in the output of the reset element jumping is fed to the lead filter whose amplification of jump leads to large control input to the system. Since low control input is preferred, the optimal sequence from a precision perspective is also the optimal sequence from the control input viewpoint.
	\subsection{Shaping Filter}
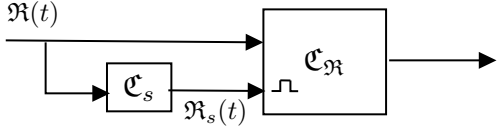
\begin{figure}[htbp]
		\centering
		\tikzset{every picture/.style={line width=0.75pt}} 
		\begin{tikzpicture}[x=0.75pt,y=0.75pt,yscale=-1,xscale=1]
		\draw   (207.94,87.35) -- (269.81,87.35) -- (269.81,140.32) -- (207.94,140.32) -- cycle ;
		\draw    (78,102.99) -- (205.94,103.98) ;
		\draw [shift={(207.94,104)}, rotate = 180.44] [fill={rgb, 255:red, 0; green, 0; blue, 0 }  ][line width=0.75]  [draw opacity=0] (8.93,-4.29) -- (0,0) -- (8.93,4.29) -- cycle    ;
		\draw   (225.47,128.64) -- (221.57,128.64) -- (221.57,123.25) ;
		\draw   (212.06,128.56) -- (216.45,128.56) -- (216.45,123.17) ;
		\draw    (216.45,123.17) -- (221.57,123.25) ;
		\draw    (98.11,104) -- (98.11,129.73) -- (128.59,129.73) ;
		\draw [shift={(130.59,129.73)}, rotate = 180] [fill={rgb, 255:red, 0; green, 0; blue, 0 }  ][line width=0.75]  [draw opacity=0] (8.93,-4.29) -- (0,0) -- (8.93,4.29) -- cycle    ;
		\draw   (129.05,114.59) -- (161.53,114.59) -- (161.53,138.81) -- (129.05,138.81) -- cycle ;
		\draw    (159.98,128.22) -- (206.97,128.54) ;
		\draw [shift={(208.97,128.56)}, rotate = 180.4] [fill={rgb, 255:red, 0; green, 0; blue, 0 }  ][line width=0.75]  [draw opacity=0] (8.93,-4.29) -- (0,0) -- (8.93,4.29) -- cycle    ;
		\draw    (271.36,113.08) -- (323.5,113.08) ;
		\draw [shift={(325.5,113.08)}, rotate = 180] [fill={rgb, 255:red, 0; green, 0; blue, 0 }  ][line width=0.75]  [draw opacity=0] (8.93,-4.29) -- (0,0) -- (8.93,4.29) -- cycle    ;
		
		\draw (238.88,113.84) node [scale=1.2]  {$\mathfrak{C}_{\mathfrak{R}}$};
		\draw (145.29,126.7) node [scale=1.2]  {$\mathfrak{C}_{s}$};
		\draw (92,90.09) node   {$\mathfrak{R}( t)$};
		\draw (184,139.41) node   {$\mathfrak{R}_{s}( t)$};
		\end{tikzpicture}
		\caption{Structure of shaping filter}
		\label{extra_control_structure}
	\end{figure}

	To attenuate the influence of noise, in sequence $No. 1$ and $No. 4$, a shaping filter $\mathfrak { C } _ { s }$ is proposed whose output is used to determine the reset instants as shown in Fig.~\ref{extra_control_structure}.
	
	This shaping filter consists of a low pass filter (LPF) and a tamed lead filter. It is represented as:
	\begin{equation}
	\mathfrak { C } _ { s }= \underbrace {\left(\frac {1} {1+\frac{s}{\omega_{f}}} \right)} _ { LPF } \underbrace {\left( \frac { 1 + \frac { s } { \omega _ { c }/a } } { 1 + \frac { s } { \omega _ { c}a } } \right)} _ { Lead }
	\label{shapefilter}
	\end{equation}
	where $\omega_f$ is the corner frequency of the LPF,$\omega_c$ is the bandwidth and $a$ is a tuning knob.
	
	The LPF plays the role of decreasing the magnitude of noise. However, the LPF also introduces phase lag into the signal used for resetting, which changes the reset instants and hence deteriorates performance. To compensate for this, a tamed lead filter is used.
	
	The phase of LPF at bandwidth can be calculated by:
	\begin{equation}
	\phi_{c}=-\tan^{-1}\Bigg(\frac{\omega_c}{\omega_f}\Bigg)
	\end{equation}
	
	To compensate for this phase change, the constant $a$ is tuned such that:
	\begin{equation}
	\tan^{-1}(a)-\tan^{-1}\Bigg(\frac{1}{a}\Bigg)=-\phi_{c}
	\label{aaa}
	\end{equation}
	
	A smaller $\omega_f$ results in greater noise attenuation, but a corresponding large value for $a$, creating a magnitude peak due to the lead filter. As a trade-off, the corner frequency of LPF is set as $\omega_f=2\omega_c$, with the corresponding $a=1.62$. 
	
\begin{figure*}[htbp]
		\centering
		\subcaptionbox{The first order DFs}
		{
			\includegraphics[width=0.4\linewidth]{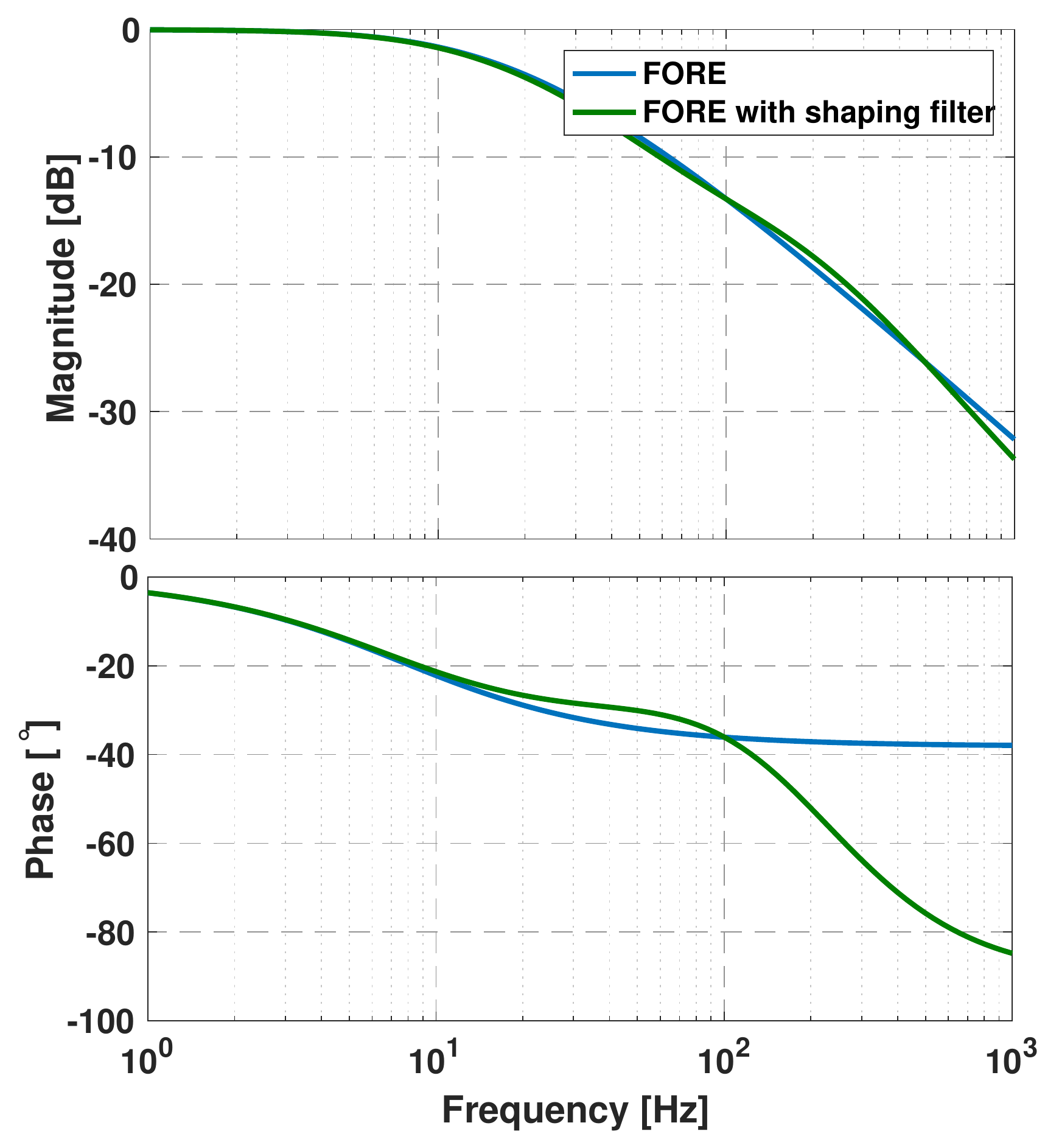}}\quad
		\subcaptionbox{The third order DFs}
		{
			\includegraphics[width=0.4\linewidth]{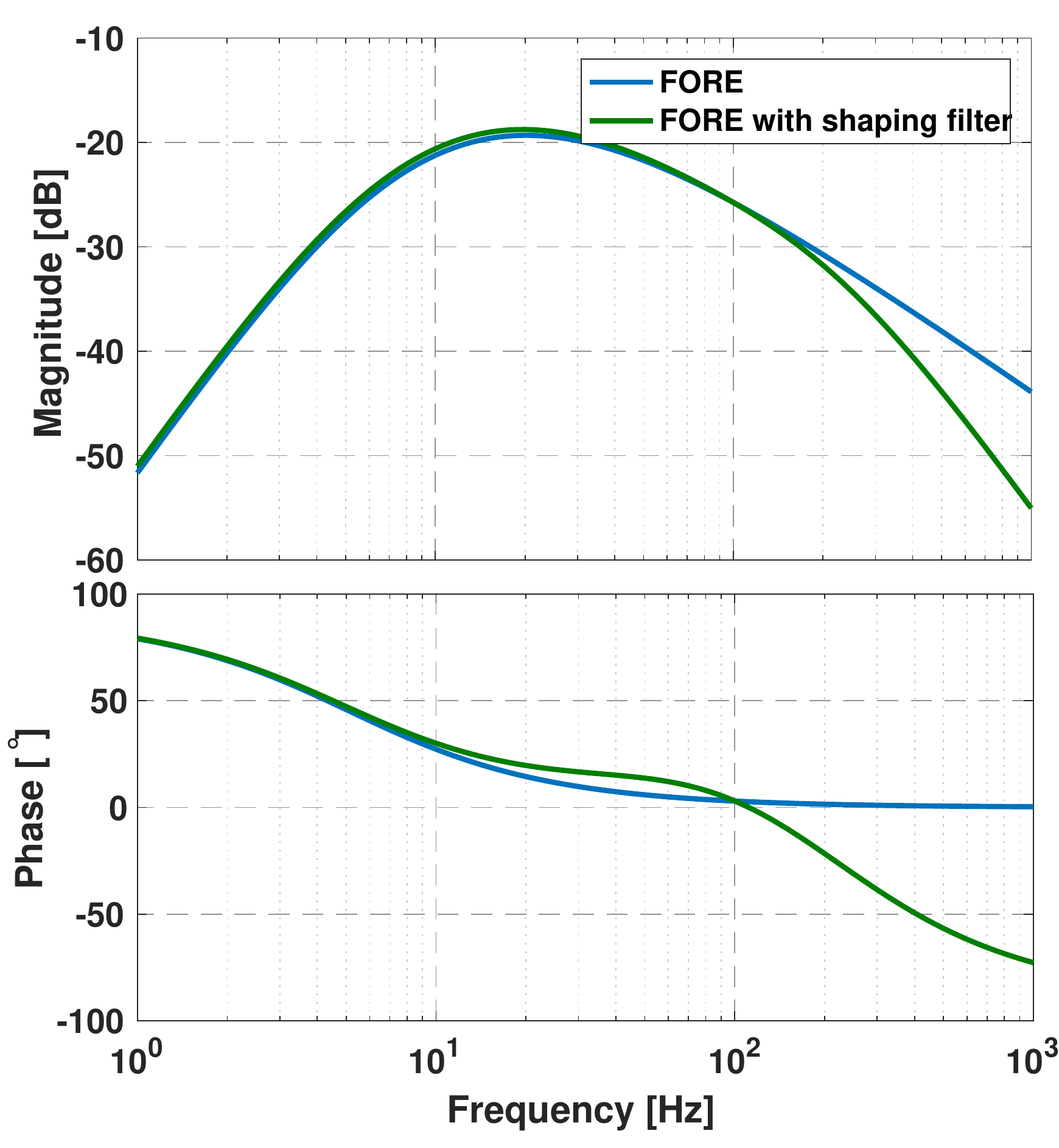}}
		\caption{The first and third order DFs of FORE and FORE with shaping filter}
		\label{DF_shaping}
	\end{figure*}

	Considering the phase of this shaping filter as $\phi(\omega)$, the HOSIDOFs of the reset element with shaping filter are re-established using a similar process in \cite{guo2009frequency} and \cite{heinen2018frequency} as:
	\begin{equation}
	\resizebox{\hsize}{!}{$
		H_n(j\omega)=\begin{cases}
		C_r(j\omega I-A_r)^{-1}(I+e^{j\phi}j\Theta_s(\omega))B_r+D_r \ \ \mbox{for}\ n=1 &   \\
		C_r(j\omega n I-A_r)^{-1}e^{j\phi}j\Theta_s(\omega)B_r   \quad \quad \quad \quad \mbox{for odd}\ n\ge2&  \\
		0  \quad \quad \quad \quad  \quad \quad \quad \quad \quad \quad \quad  \quad \quad \quad \quad  \quad \ \mbox{for even}\ n\ge2 & 
		\end{cases}$}
	\end{equation}
	where
	\[\Theta_s=\Theta_\rho(\frac{-A_r \sin\phi+\omega \cos\phi I}{\omega})\]
	The first and third order DF of the traditional FORE and the FORE with shaping filter are shown in Fig.~\ref{DF_shaping}. It can be seen that the shaping filter does not change the DF significantly, but the magnitude of the third order harmonic is reduced after $\omega_f$.
	\begin{figure}[htbp]
		\centering{\includegraphics[width=\columnwidth]{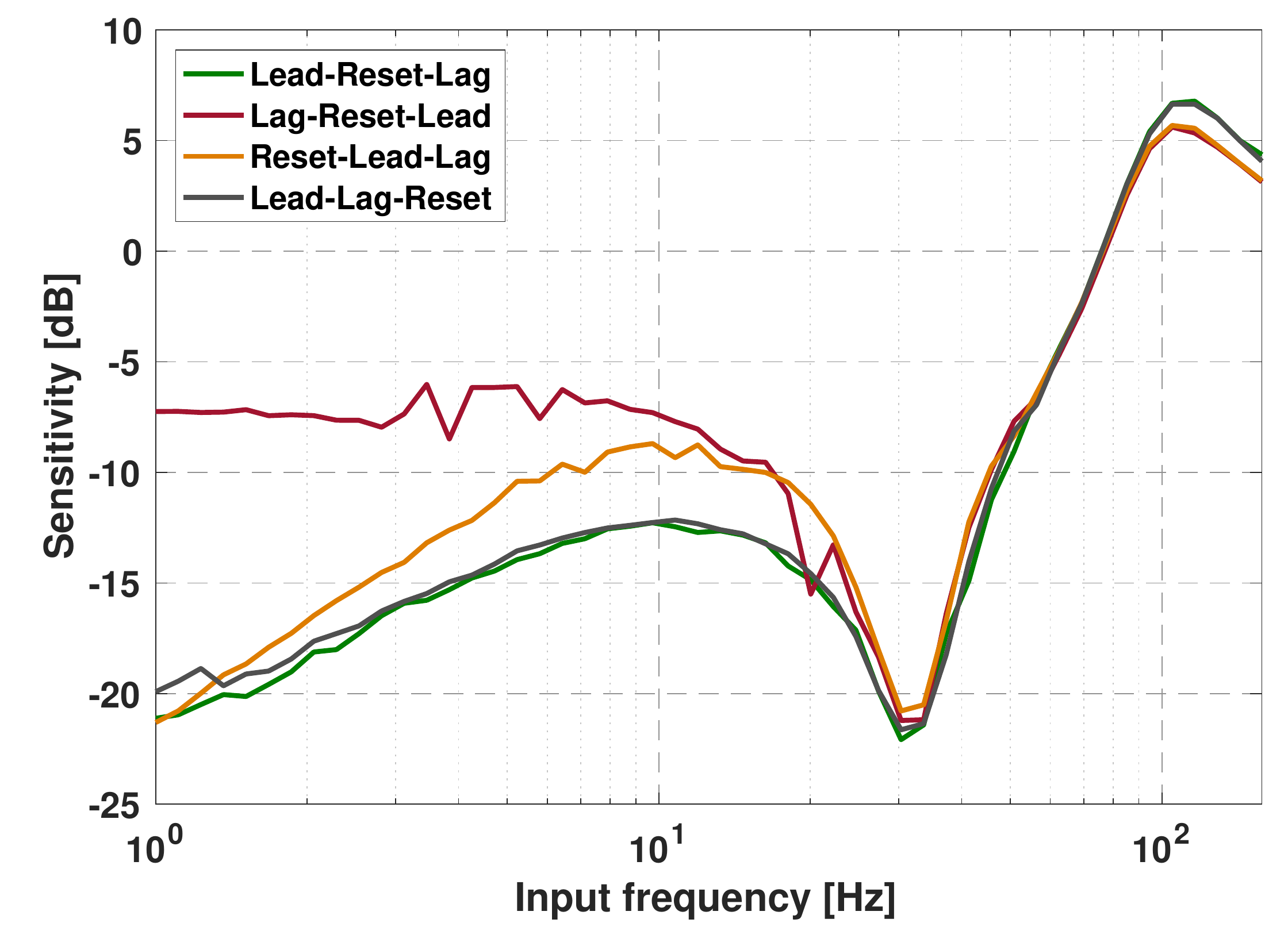}}
		\caption{Sensitivity function with $3\%$ noise when shaping filter is applied on sequence 1 and 4}
		\label{S with additional control2}
	\end{figure}
	
	The pseudo sensitivity  $S _ { \partial } ( \omega )$ is obtained with the use of shaping filter for sequences $No.1$ and $No.4$ and is shown in Fig.~\ref{S with additional control2} with $3\%$ noise added. This shaping filter drastically reduces the effect of noise and improves tracking performance. Although the performance deteriorates slightly around the bandwidth, this will not affect the tracking performance of trajectory signals in reality, where high-frequency components are often pre-filtered out \cite{lambrechts2005trajectory}. Simulation performance for noise levels larger than $3\%$ showed poor performance for the chosen shaping filter and hence are not shown. For larger levels of noise, shaping filter with smaller $\omega_f$ needs to be used. However, the $3\%$ noise level is already quite large for several precision positioning applications and hence this technique can be successfully used in practice.
	
	In summary, the sequence $No.1$ has the optimal sequence for tracking performance for noise signal up-to $1\%$ amplitude compared to the reference. For larger noise levels, a shaping filter can be used to attenuate effects of noise in the performance. Furthermore, the sequence $No.1$ has the minimum control input among all possible sequences.
	
	\subsection{Step Response}
\begin{figure}[htbp]
		\centering{\includegraphics[width=\columnwidth]{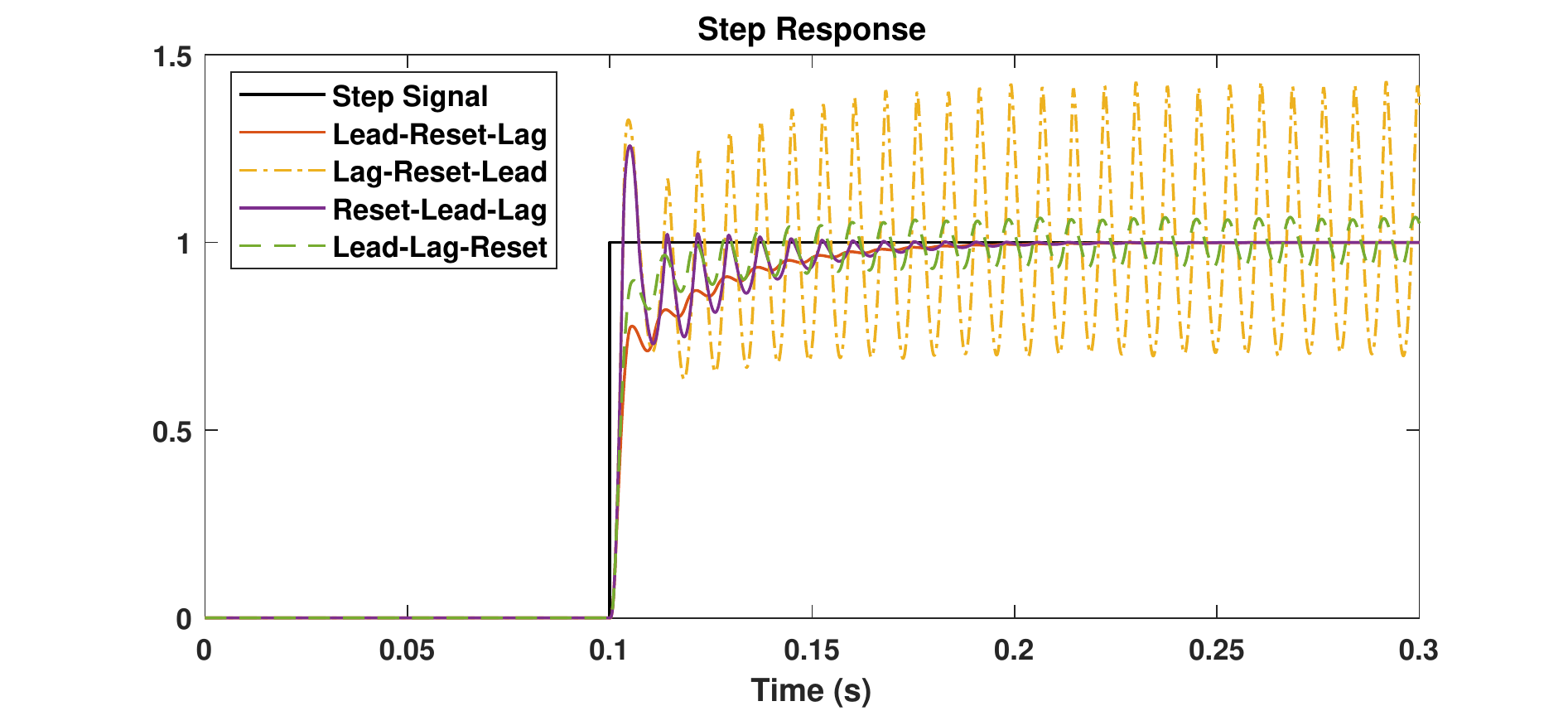}}
		\caption{Step responses of different sequences}
		\label{stepresponse}
	\end{figure}

	The step responses of different sequences are compared in Fig.~\ref{stepresponse}. It can be seen that steady state error is seen when integrator (lag filter) is in front of the reset element (the system is not asymptotically stable). Also, overshoot occurs when differentiator (lead) is located after the reset element. Although putting the lead filter after the reset element has less rise time than putting it before the reset element, both sequences have the same settling time. From the time domain perspective, Lead-Reset-Lag (sequence $No. 1$) is still the optimal sequence.
	
	\section{EXPERIMENTAL VALIDATION}\label{five}
	To validate the simulation results, a series of experiments are conducted for all the four sequences without any shaping filters. The maximum error values along with the maximum control input values are obtained for five different frequencies of reference input. Further, to avoid problems due to control input saturation, different amplitudes are chosen for the sinusoidal reference signals at different frequencies as given in TABLE~\ref{reference magnitude}. The amplitude of the noise in the system is found to be ($100\text{nm})$ for the experiments. The maximum steady-state error signal ($\max(|e(t)|)$) and maximum steady-state digital control input are recorded in TABLE~\ref{1experiment} and TABLE~\ref{2experiment}  respectively.
	
	\begin{table}[htbp]
		\caption{Magnitude of sinusoidal reference and the level of noise}
		\centering
		\begin{tabular}{|c|c|c|}
			\hline
			\multicolumn{2}{|c|}{Reference signal}&level of noise\\
			\hline
			Frequency(Hz) & Magnitude $(0.1\mu m)$ & Percentage \\
			\hline
			1 & 100 & 1\%   \\
			\hline      
			5 & 120 & 0.83\% \\ 
			\hline
			10 & 120 & 0.83\% \\
			\hline       
			15 & 150 & 0.67\% \\
			\hline
			20 & 200 & 0.5\% \\
			\hline
		\end{tabular}
		\label{reference magnitude}
	\end{table}

	\begin{table}[htbp]
		\caption{Maximum steady-state error of four different sequences}
		\centering
		\begin{tabular}{|c|c|c|c|c|}
			\hline
			Reference & \multicolumn{4}{|c|}{$\max(|e(t)|$) (0.1$\mu m$)}\\
			\hline  
			(Hz)  &    No.1 & No.2& No.3& No.4\\
			\hline
			1  & 15 & 73 & 14 & 15 \\
			\hline
			5  & 37  & 78  & 56  & 40 \\
			\hline
			10  & 42  & 73  & 67  & 48\\
			\hline
			15  & 54 & 84  & 86 & 54 \\
			\hline
			20  & 55  & 97  & 86  & 55 \\
			\hline
		\end{tabular}
		\label{1experiment}
	\end{table}
	
	\begin{table}[htbp]
		\caption{Maximum steady-state control input of four different sequences}
		\centering
		\begin{tabular}{|c|c|c|c|c|}
			\hline
			Reference & \multicolumn{4}{|c|}{Digital control input (count)}\\
			\hline  
			(Hz)  &    No.1 & No.2& No.3& No.4\\
			\hline
			1   & 486  & 26173 & 3103 & 1222\\
			\hline
			5  &  884  & 26806 & 12785  & 1513\\
			\hline
			10   & 941  & 24476  & 16972  & 1306\\
			\hline
			15  & 1364 & 25718 & 19539  & 1473\\
			\hline
			20   & 1677  & 27541  & 22038  & 1471\\
			\hline
		\end{tabular}
		\label{2experiment}
	\end{table}
	
	The results in the tables validate the theory and simulation results as sequence $No.1$ provides the lowest error at almost all tested frequencies except at 1Hz, where $No.3$ has a lower error. This is consistent with the simulation results since the noise level being 1$\%$ of reference amplitude at this frequency results in performance deterioration. In addition, at $15Hz$ and $20Hz$ which are both more than $\omega_i$, the effect of the integrator is vanished ($\dfrac{\omega_i}{s}|_{\omega>\omega_i}\approx0$). Consequently, the sequences $No.1$ and $No.4$ have the same performance.
	
	To check the effect of noise at low frequencies and the effect of shaping filter in overcoming this problem, a different set of experiments is conducted at 1Hz with $3\%$ noise. Since sequence $No. 2$ is the worst sequence in terms of both tracking performance and control input as seen in Tables. \ref{1experiment} and \ref{2experiment}, this sequence is not tested for and only the performance of the other three sequences are compared in TABLE~\ref{3experiment}. Without the shaping filter, the performance of $No.1$ and $No.4$ significantly deteriorates, while the performance of $No.3$ does not change a lot with an increase in noise levels. When the shaping filter is applied, the performances of $No.1$ and $No.4$ are improved significantly which means that the effect of noise is effectively suppressed. The efficacy of the shaping filter is hence verified in practice.
	\begin{table}[htbp]
		\caption{Influence of shaping filter on maximum steady-state error}
		\centering
		\begin{tabular}{|c|c|c|c|c|}
			\hline
			\multirow{2}*{Configuration}& \multirow{2}*{level of noise}& \multicolumn{3}{|c|}{$\max(|e(t)|$) (0.1$\mu m$)}\\
			\cline{3-5} 
			~ &~ & No.1& No.3& No.4\\
			\hline
			without shaping filter& 1\% & 15 & 14 & 15 \\
			\hline
			without shaping filter& 3\% & 49 & 20 & 32 \\
			\hline
			with shaping filter& 3\% & 19 & 19 & 21 \\
			\hline
		\end{tabular}
		\label{3experiment}
	\end{table}

	\section{CONCLUSIONS}\label{six}
	This paper has proposed an optimized strategy for the sequence of controller parts when a reset element is used. Firstly, the frequency responses of the different sequences were investigated by considering high order harmonics using HOSIDF theory. The optimal sequence is hypothesized to be the one in which the magnitude of high order harmonics is minimum. Next, the closed-loop performances of a high-tech positioning stage with PI+CgLp controller were analyzed in both simulation and experiment for different sequences of controller parts. The results illustrate that when the magnitude of noise within the system is smaller than $1\%$ of the reference signal, it is safe to say that the suggested sequence has the best performance. Otherwise, the performance of the suggested sequence will deteriorate at low frequencies. In this case, a shaping filter is proposed to deal with the problem. It is revealed that this shaping filter attenuates the influence of noise successfully and allows the suggested sequence to provide the best tracking performance with up to $3\%$ noise. In addition, the suggested sequence also has the smallest control input, which provides greater flexibility for actuator choice/design.
	
	These results can facilitate the use of reset controllers in a broad range of applications in high-tech industry. Application of this approach for other kinds of nonlinear controllers for improved performances is a promising topic for investigation in the future.
	
	\addtolength{\textheight}{-12cm}   
	


	\bibliographystyle{IEEEtran}
	\bibliography{phd}
\end{document}